\documentclass[aps,prd,nofootinbib,twocolumn,longbibliography]{revtex4-1}
\usepackage[english]{babel}
\usepackage[autostyle, english = british]{csquotes}
\usepackage[CJKbookmarks, pdftex, bookmarksnumbered, bookmarksopen, colorlinks, citecolor=blue, linkcolor=blue]{hyperref}
\usepackage{amsmath,amssymb}
\usepackage{graphicx}
\usepackage{enumitem}
\setlist[enumerate]{topsep=0pt,parsep=-1mm,leftmargin=5mm,}
\def\be{\begin{equation}}
\def\ee{\end{equation}}

\newcommand*{\diff}{\mathop{}\!\mathrm{d}}

\begin{document}

\title{Planck stars, White Holes, Remnants and Planck-mass quasi-particles\\[1mm]
{\normalsize The quantum gravity phase in black holes'  evolution and its manifestations}
% {\centerline{\rm DRAFT}}
}

\author{Carlo Rovelli${}^{abcd}$, 
%Farshid Soltani${}^a$, 
Francesca Vidotto${}^{def}$}

\affiliation{${}^a$ Aix-Marseille University, Universit\'e de Toulon, CPT-CNRS, 13288 Marseille, France.}
\affiliation{${}^b$ Perimeter Institute, N2L2Y5 Waterloo, Ontario, Canada} 
\affiliation{${}^c$ Santa Fe Institute, 87501 Santa Fe, New Mexico, USA} 
\affiliation{${}^d$ {Philosophy Department and Rotman Institute,  Western University, N6A5B7 London, Ontario, Canada}}
\affiliation{${}^e$ Physics and Astronomy Department, Western University, N6A5B7 London, Ontario, Canada}
\affiliation{${}^f$ Instituto de Estructura de la Materia, IEM-CSIC, Serrano 121, 28006 Madrid, Spain}
        
\begin{abstract} 
\noindent 
This is a review  of some recent developments on quantum gravity aspects of black hole physics. In particular, we focus on a scenario leading to the prediction of the existence of a Planck-mass quasi-stable object, that could form a component of dark matter.

\end{abstract}

\maketitle

\tableofcontents

\section{Introduction}
\noindent  Quantum gravity is a theory with a mass scale: $m_P=\sqrt{c\hbar/G}$, a fraction of milligram. This is a small scale in astrophysics and a large scale in high-energy physics.  It is reasonable to study the possibility that the spectrum of the theory could include a stable or semi-stable non-perturbative object at this scale: a Planck-mass quasi-particle.   Recent developments in classical general relativity (GR) and in LQG (LQG) add plausibility to this possibility and suggest the existence of a semi-stable object at the mass
\be
m=\sqrt{\frac{\sqrt{3}\gamma c \hbar}{4G}}\sim 14\sqrt{\gamma}\;\mu g
\label{remmass}
\ee
where $\gamma$ is the Barbero-Immizi parameter, a free parameter in LQG, akin to the  $\theta_{QCD}$ angle in  quantum chromodynamics, and presumed of order unity.  These developments emerged from studying the dynamics of black holes.   

We expect black holes to evolve into spacetime regions dominated by strong quantum gravity effects. A number of recent lines of research have studied these regions, and found an interesting  scenario for the full evolution of a black hole \cite{Rovelli2018h}.  Different ingredients  contribute to this scenario. These include a new solution of the Einstein equations \cite{Haggard2014} showing that a trapping horizon can evolve into an anti-trapping one, a better understanding of the interior of white holes and black holes \cite{Christodoulou2015}, and numerous applications of a variety of LQG techniques --canonical, covariant, and numerical-- to describe the genuinely non-perturbative regions \cite{Modesto:2004xx,Ashtekar:2005cj,Rovelli2014ps,Tavakoli2015,Bianchi2018e,Lewandowski2022,Husain2022a,Han:2023wxg,Frisoni2023,Han2024,Sobrinho2023,Borges2024}.   

Three aspects of this scenario are particularly appealing.  (a) It offers a natural solution to the black hole information `paradox'.   (b) It is in principle, and perhaps even in practice, directly testable \cite{Perez2023}. And (c) it provides a natural candidate for dark matter that does not require any new physics (such as new fields, particles, or modifications of the field equations): just general relativity and its quantum properties \cite{Rovelli2018f}.  This scenario is compatible with and possibly corroborating the general idea that primordial black holes could play a key role for explaining dark matter, see  \cite{
carr_primordial_2020,green_primordial_2021}.
%Carr:2020xqk,Green:2020jor}
%Banks2021,Banks2020}

This scenario combines  distinct quantum phenomena happening in different spacetime regions. It also includes dissipative as well as non-dissipative aspects. The analysis of its various  aspects  employs different approximations and truncations for treating  different phenomena.  Because of this complexity, it can only be addressed `\`a la Fermi', estimating the relevance and the import of the various physical effects, rather than within a single mathematical-physics idealization.  This complexity motivates the present review paper, which brings together several ingredients for this scenario, scattered in the literature. The only truly original part of this review is Section \ref{spread}, which present an argument supporting the idea that a Planck-mass quasi-particle produced as a black hole remnant could be understood as a superposition of a black and a white horizon.  

We start with a quick sketch of the scenario (Section \ref{sketch}) and an analysis of the regions where classical GR is unreliable (Section \ref{regions}). Then we break the presentation into two parts: a first part  (Section \ref{I}) where we discuss the non-dissipative (time-reversal symmetric) aspects of the global dynamics of a black hole, and a second part  (Section \ref{II}) where we take the dissipative (time oriented) effects into account.  

In the first part, we discuss the bounce of a collapsing matter distribution (Section \ref{PlanckStars}), namely the `Planck Star' phenomenon, and the external metric of the global black hole dynamics  (Section \ref{transition}).  A side section discusses the counter-intuitive classical geometry of white holes, because this play a role in global picture  (Section \ref{WhiteHoles}).  This part ends presenting the current state of the calculations of the black-to-white transition probability  \cite{Christodoulou2018}.   

In the second part, based on these calculations, we discuss the relevant temporal regimes, the black hole information paradox (Section \ref{InformationParadox}), and the structure of and stability of the remnants (Section \ref{Remnants}). The last part (Section \ref{DarkMatter})  focuses on the possibility of observations supporting the scenario described.
Unless indicated, we use Planck units $c=G=\hbar=k=1$.

\subsection{A sketch of the scenario}\label{sketch}

Perhaps the most interesting discovery of the last decades is the abundance and variety of the astrophysical objects we now call black holes.  Current direct and indirect observations of these  objects are all well accounted for by classical GR. But classical GR is insufficient to account for the phenomena that happen in their high-curvature regions, where genuine quantum gravity effects are most likely non-negligible. Therefore GR does not provide a reliable \emph{global} picture of the dynamics of these objects. 

In a black hole produced by  a collapsing matter distribution there are three distinct spatiotemporal regions where we expect quantum gravity effects to dominate: (i) the region when the matter distribution reaches Planck density;  (ii) the region, outside the matter distribution but inside the horizon, when Planck curvature is reached; (iii) the region \emph{outside} the horizon, where the Planck curvature is reached because of the shrinking of the horizon due to the Hawking radiation. These events are spacelike with respect to one another.  Each region needs to be treated separately and understood on its own terms.  But all of them can be described using LQG as a guide, utilizing tools and results from various branches of LQG, in particular the spinfoam amplitudes of covariant LQG \cite{Rovelli2011c,Rovelli2015}, Loop Quantum Cosmology \cite{Ashtekar2015a,agullo_loop_2016} and numerical calculations \cite{dona_how-compute_2022,han_mathematica_2024}.  

The intuition underpinning the scenario that emerges from combining these results  \cite{Bianchi2018e} is that quantum-gravitational pressure can stop gravitational collapse and cause a bounce \cite{Rovelli2014ps}, permitting the entire content of the black hole to eventually dissipate \cite{Ashtekar:2005cj}.   This is the same phenomenon as in loop quantum cosmology, that indicates that the dominant quantum effect at high density is a quantum pressure sufficient to counterbalance weight and reverse collapse.  

This quantum gravitational pressure stops gravitational collapse when the energy density becomes of Planckian order, yielding a new phase in the life of gravitationally collapsed objects. %\cite{Dai2010}
At the bounce itself, the matter distribution has maximal density and is called a ``Planck star" \cite{Rovelli2014ps}. After this, the collapsing object bounces out.  In the meanwhile, the geometry outside the star, including the horizon, also bounces tunnelling \cite{Dona2024} from a black to a white hole geometry.    This process can be  short in proper time, as well as, due to the huge gravitational time dilation, very long for an external observer.

The process may  seem incompatible with Birkhoff's theorem (in the spherically symmetric case), and this may be the reason for which the possibility for it to happen was long been missed.  As detailed later on, there is no incompatibility, because the Birkhoff's  theorem is local but not global, hence  
%Birkhoff's theorem states that any spherical symmetric solution to the vacuum Einstein's  equation is isometric to the Kruskal solution, and the Kruskal solution does not permit a black hole to evolve into a white hole. But it was realized in that there is an exact solution of the Einstein equations,  which is locally --but not globally-- isometric to the Kruskal spacetime, which can surround a black hole that tunnels into a white hole, where Einstein's equation are only violated within the small high curvature spatiotemporal region.  
quantum tunnelling in a compact region can circumvent the theorem \cite{Haggard2014} and 
%A local quantum violation of classical GR in a compact region, namely a conventional tunnelling effect,
permit a black hole to evolve into an anti-trapped region, that is, a white hole. 
  This scenario has been  explored  in \cite{DeLorenzo2016,Christodoulou2018,Bianchi2018e, DAmbrosio2018, Rovelli2018a,Fazzini2023}.

If the transition happens at the end of the evaporation, when quantum gravity becomes relevant also outside the horizon, and where quantum gravitational transition amplitudes indicate its probability to approach unity, most of the energy of the black hole has already been radiated away into Hawking radiation, which is a dissipative phenomenon that breaks time-reversal invariance. Therefore, even if the transition itself can be described as an elastic non-dissipative phenomenon, the overall life of a black hole is very far from being time-reversal symmetric, and we may expect the white hole emerging from the transition to be a remnant with mass at the Planck scale. 

The internal geometry of these objects is a remarkably precise realization, fully consistent with classical GR, of the `cornucopia' intuition considered in the 1980's and later abandoned on the  basis of arguments that (as we shall discuss below) are no longer cogent \cite{Banks1992b,Banks1993}.   

There is a well-known instability affecting eternal and macroscopic white holes: they can easily fall into a black hole as in the process described by the classical Kruskal geometry.  On the  other hand, there are arguments indicating that a  Plank-mass white hole can be stabilized by quantum gravity \cite{Rovelli2018f}, by shifting it into a quantum superposition of black and white geometries.  As we show below, the key of this stabilization is the LQG area gap, because of which there is likely no black (or white) hole with a mass smaller than the Planck mass. 

Hence the scenario predicts the existence of quasi-stable remnants of a known mass, determined by the LQG area gap.  A large number of these, of primordial or possibly even pre-big-bang origin, would behave precisely as dark matter.  The possibility of direct detection of these objects has been explored in  \cite{Perez2023} and we review it below.  

Once quantum gravity phenomena are taken into account, the notion itself of horizon is necessarily different than the one relevant for the classical theory.  \emph{Event} horizons lose their relevance, because they are likely not to form at all. The Hawking radiation drags the dynamical evolution of the horizon into regions where the Einstein equations do not hold anymore.  Accordingly, the discussion on the so called black-hole information paradox needs to be reconsidered. 

 Even more dramatically, the traditional \emph{definition} of a black hole as an object characterized by an \emph{event} horizon becomes misleading: in the light of quantum gravity,  the horizon of the realistic astrophysical black holes is likely \emph{not} to be an \emph{event} horizon, because while the Kerr geometry and its perturbations used in modelling astrophysical objects may be appropriate, its future time evolution (which determine whether the horizon is an event horizon) is not the one determined by the Einstein equations.   Of course the local properties of the horizon (described as apparent \cite{Hawking_Ellis},  trapping  \cite{Hayward1994}, isolated  \cite{Ashtekar2000}, dynamical \cite{Ashtekar:2004cn}, and similar, see this last reference for a review) remain the same. 

While the scenario illustrated takes fully into account quantum gravitational phenomena in the high curvature region, it assumes that classical GR and quantum field theory on curved spacetimes are good approximations outside these regions.  Contrary to what sometimes claimed, GR and quantum field theory taken together, yield no inconsistencies in the entire region where quantum gravity can be neglected.  As argued in detail below, the common claim that inconsistencies appear already at the Page time depends on strong versions of the holographic hypothesis \cite{Rovelli2017a}, or from confusion between ADM mass and Bondi mass, between thermodynamic and von Neumann entropies, or between event and dynamical horizons. On the contrary, the scenario considered here is only based on conventional GR and conventional quantum ideas.

 Yet, this scenario implies a number or remarkable new phenomena: (i)  Due to the huge time dilation in a black hole, the process can last micro-seconds in local proper time, but billions of light-years observed from the outside; (ii) the internal volume of the hole can remain huge even as the size of the horizon shrinks; (iii) the star's bounce volume is much larger than Planckian, because the onset of quantum-gravity effects is governed by density, not size; (iv) the interior of an evaporating hole can keep memory of the initial state, without information loss; (v) information can follow a different path from energy, and be slowly released by the remnants, while most of the energy was previously lost into the Hawking radiation. All this shows that unpalatable phenomena like `firewalls' \cite{Almheiri2013a} are not to be expected on the basis of GR and quantum theory alone, unless one adopts arbitrary holographic assumptions that are at best string-motivated only.

\subsection{The domain of validity of classical gravity} \label{regions}

 \begin{figure}[t]
    \centering
    \includegraphics[width=4cm]{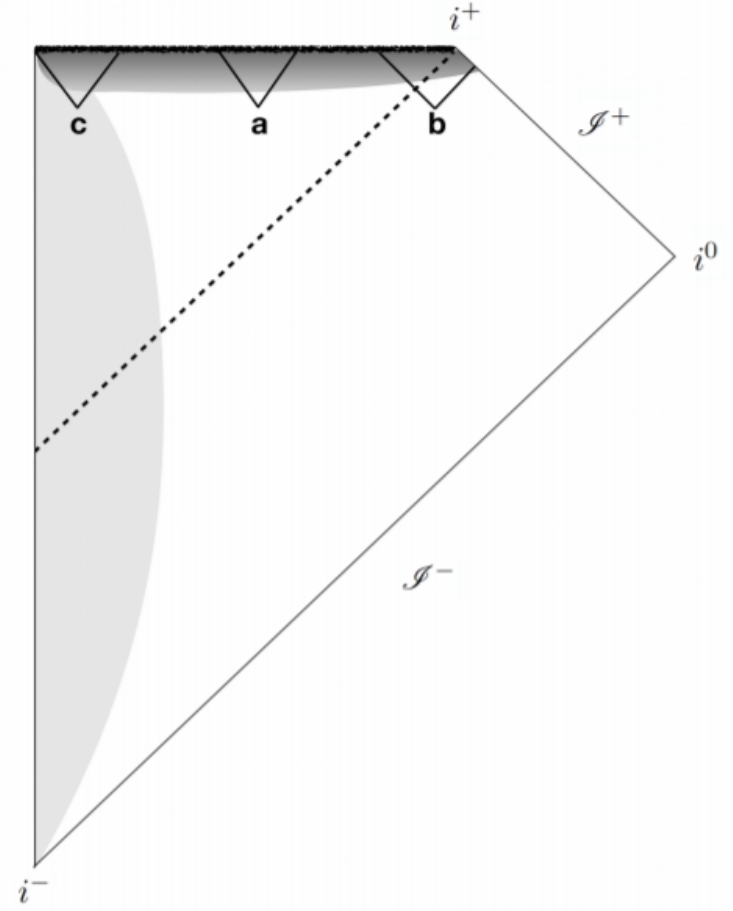}
    \caption{The Carter-Penrose diagram of a Schwarzshild black hole until the onset of quantum gravity. The light grey region is the collapsing star. The dark grey region is where quantum gravity becomes relevant.}
    \label{SchBH}
\end{figure}

What happens inside a black hole is dynamical: it cannot be understood in terms of any static or stationary model.
The Carter-Penrose diagram of a (classical) Schwarzschild black hole created from a gravitational collapse is depicted on Figure~\ref{SchBH}. The dark grey region is where the classical theory becomes unreliable, due to quantum gravitational effects. We expect this to happen when the curvature becomes Planckian, for instance when the Kretschmann scalar 
\begin{equation}
    K^2 = R_{\alpha \beta \gamma \delta}R^{\alpha \beta \gamma \delta} = 48\frac{M^2}{r^6}
    \label{kretscalar}
\end{equation}
becomes of order 1 in natural units. Here  $M$ is the black hole mass and $r$ is the Schwarzschild radius.  This happens before the $r=0$ singularity inside the black hole. Notice that if $M$ is macroscopic, it happens at a radius of order  $r\sim M^{\frac13}$ much larger than the Planck radius ($r\sim 1$).

Importantly the surface where this happens is space-like.  

Just outside the horizon, where $r\sim 2M$, $K \sim \frac{M}{r^3}\sim \frac{1}{M^2}$. The Hawking evaporation steadily decreases the mass $M$ of an isolated black hole, bringing $K$ up to Planckian values, hence \emph{the quantum region extends  outside the horizon}.  As we will see later on, results in \cite{Christodoulou2018,Han2024} indicate that the transition probability $P$ from black holes to white holes is proportional to 
\begin{equation}
    P \sim e^{-M^2}.
    \label{probatuunel}
\end{equation}
Thus becoming dominant at the end of the evaporation, where $M\sim 1$.  

The possibility  that quantum effects could appear earlier, at a (retarded) time of order $M^2$ after the collapse,  has been considered in \cite{Haggard2014, Haggard20162} and in \cite{Husain2022,Husain2022a,Fazzini2023} by considering the interplay with the collapsing matter. The phenomenology this could give rise to has been explored extensively in \cite{Barrau2014b,Barrau2014,Barrau2016c,Vidotto2018,Bull2018,Barrau2018b}. Here we will not consider this hypothesis.    

The region of the onset of quantum gravitational phenomena can thus be organized into three sub-regions \cite{Dambrosio2021} (see Figure \ref{SchBH}
):
\begin{itemize}
\setlength\itemsep{.5mm}
\item Region \textbf{a} :  The region which is neither directly causally connected to the horizon nor to the collapsing matter distribution (it is \emph{amid} them).
\item Region \textbf{b} : The region in the vicinity of the horizon (\emph{the boundary}). 
  \item Region \textbf{c} : The collapsing matter distribution region (\emph{the center}). 
  \end{itemize}
The onset of the phenomena in these three regions are causally disconnected, as they are separated by space-like distance. For a black hole of (initial) mass $M$, this distance is of the order  \cite{Dambrosio2021} 
\begin{equation}
    L \simeq {M}^\frac{10}{3},
    \label{interior}
\end{equation}
which is colossal for a macroscopic black hole.  Notice that the locus of the onset of quantum gravity is not `at a point', but rather `at a time'.    It is not `somewhere' but rather, `at some time', which, in relativistic terms, means in a set of space-like related regions. 

The fact that the various onsets of quantum gravity are spatially related implies by locality that what happens in a region of this locus is causally disconnected, hence independent from what happens in other regions.  To imagine that quantum gravity could violate this macroscopic causal disconnection is in gross contradiction with what we know about reality, not supported or suggest by anything we know, as far as we can see. 
 To understand the physics of the end of a black hole, we have to understand the quantum evolution of each of these three regions independently.

\section{Non dissipative aspects of the transition}\label{I}

\subsection{Planck stars}\label{PlanckStars}

Let us start by studying what happens in the interior of a collapsing star, after it enters its horizon, when it reaches Planckian density. The simplest modelling of a collapsing matter distribution 
is provided by the Oppenheimer-Snyder model \cite{Oppenheimer:1939ue}: a spherically symmetric pressure-less homogeneous matter distribution 
free-falling under its own weight.  Assuming the matter distribution to start at rest in the distant past, the metric inside a such a matter distribution 
distribution can be written in co-moving and proper time coordinates $(T,R)$ as 
\be
\diff s^2=-\diff T^2+a^2(T)(\diff R^2+R^2\diff\Omega^2)\, ,
\label{star}
\ee
where the $T=constant$ slices define the homogeneity foliation, $\diff \Omega^2$ is the metric of the unit 2-sphere, $R\in[0,R_{boundary}]$ and $a(T)$ is known as the scale factor. The radial co-moving coordinate of the boundary of the matter distribution 
distribution can be chosen to be $R_{boundary}=1$ without loss of generality. The uniform density of the matter distribution 
distribution is then $\rho=m/\frac43\pi a^3$, where $m$ is the total mass.   Inserting this metric in the Einstein field equations gives the Friedmann equation for $a(t)$:
\be
\frac{\dot{a}^2}{a^2}=\frac{8 \pi}{3} \rho\,,
\label{Fried}
\ee
where the over-dot means differentiation with respect to $T$, which coincides with the proper time.  Eq.~\eqref{Fried} can be solved:
\be
a(T)= \bigg(\frac{9m (T-T_0)^2}{2 R^3_{\text{star}}} \bigg)^{1/3}  .
\label{aclassic}
\ee
Without loss of generality we can take the time at which the matter distribution 
distribution collapses to zero physical radius to be $T=0$.

\begin{figure}[t]
\begin{center}
\includegraphics[width=5cm]{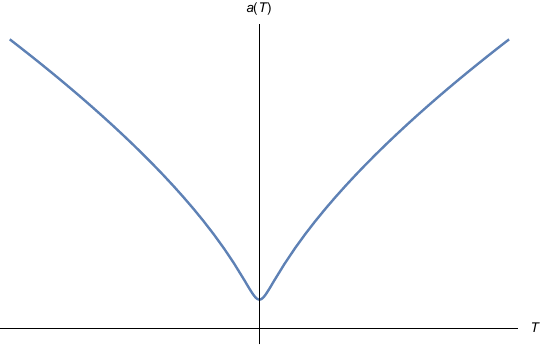}
\caption{The scale factor $a(T)$ in eq.~\eqref{aq} that gives the standard LQC bounce.}
\label{bounce2}
\end{center}
\end{figure}

How does quantum gravity affects this dynamics? A major result in loop quantum cosmology~\cite{Ashtekar2006,Ashtekar:2006uz,Agullo2013a,Kelly:2020lec} is that the Friedmann equation for the scale factor, Eq.~\eqref{Fried}, is modified by quantum gravity effect into 
\be
\frac{\dot{a}^2}{a^2}=\frac{8 \pi}{3} \rho\, \Big( 1-\frac{\rho}{\rho_c}\Big)\,,
\label{Friedq}
\ee
where $\rho_{c}= \sqrt{3}c^2/(32 \pi^2 \gamma^3 \hbar G^2)\sim c^2/\hbar G^2$, $\gamma$ being the Barbero-Immirzi parameter, is the critical density \cite{Ashtekar:2011ni}, a constant with the dimension of a density and Planckian value. This equation can be integrated to give
\be
a(T)= \bigg(\frac{9m T^2 +A m}{2 R^3_{\text{star}}} \bigg)^{1/3},
\label{aq}
\ee
where $A=3/(2\pi\rho_c)$ is a parameter of Planckian value. In units where $G=c=1$, the constant is $A\sim \hbar \sim m_{Pl}^2$.  Notice that $a(T)$ is positive for the whole range $T\in [-\infty, \infty]$: it decreases for $T<0$, reaches a minimum $a_0=\sqrt[3]{Am/2}$ for $T=0$ and then increases for $T>0$.  See  Fig.~\ref{bounce2}.  This is the characteristic bounce of loop quantum cosmology. This result shows that the line element in eq.~\eqref{star} is well defined everywhere, without singular collapse: the star reaches a maximal density, where it is called a `Planck star', and then bounces and expands.  

Of course the time-reversal symmetric bounce is an approximation, because dissipative effects break this symmetry. The Hawking radiation generates an ingoing flux of negative energy  that is likely to compensate the energy of the star, in the interior of the hole.    We will address these phenomena later on, when discussing the dissipative aspects of the dynamics of the black hole. 

To get a feeling for the physics of the bounce, we can rewrite \eqref{Friedq} in the form  
\be
\dot a^2=\frac{2m}{a}-\frac{Am^2}{a^4}.
\label{eqaT}
\ee
The coordinate $T$ is the proper time along the comoving worldlines, hence it is also the proper time on the boundary of the star. This means that Eq.~(\ref{eqaT}) gives also the evolution of the physical radius $r_b(T)=a(T)$ of the matter distribution%star
in its own proper time, hence
\be
\dot r_b^2=\frac{2m}{r_b}-\frac{Am^2}{r_b^4}\,. 
\label{boundaryint}
\ee
This shows that a mass element on the boundary of the matter distribution%star
falling in its own proper time feels a potential that is the Newtonian one, precisely as in classical GR, but corrected by a repulsive quantum term proportional to the inverse of the 4th power of the radius.   This is the short scale \emph{repulsive} force due to the quantum pressure, that grows stronger at smaller radius. 

Is this dynamics compatible with the dynamics of the surrounding geometry? 

\subsection{Black-to-white transition}\label{transition}

In the classical Oppenheimer-Snyder model, the geometry of the collapsing matter%star
is compatible with a surrounding Schwarzschild geometry. That is, matching conditions between the two geometries are satisfied on the boundary of the matter distribution. To confirm this, consider the surface bounding the collapsing matter as a free falling spherically symmetric shell in a Schwarzschild metric
\be
\diff s^2=-F(r)dt^2+dr^2/F(r)+r^2\diff\Omega^2\, ,
\label{Schwarzschild}
\ee
where $F(r)=1-2m/r$.  The shell follows a radial time-like geodetic and therefore satisfies 
\be
-1=-F(r)\dot t^2 +\dot r^2/ F(r) 
\ee
The conservation law associated to the time-like Killing symmetry gives $F(r)\dot t=E$ where $E$ is a constant that can be taken equal to unit if the shell starts at rest at infinite radius. Hence $-F(r)=-E^2 +\dot r^2$, that is
\be
\dot r^2=\frac{2m}{r}.
\ee
A comparison with \eqref{boundaryint} shows that this is exactly the change of the Schwarzschild radius in proper time of the boundary of the star, in the classical ($A$=0) case.   

But this suggest immediately the form of a metric which is compatible with the bouncing star, in the quantum case: it is again \eqref{Schwarzschild} but with 
\be
F(r)=1-\frac{2m}{r}+\frac{Am}{r^4}.
\label{F}
\ee
This is an interesting metric. (For earlier attempts to write the black hole metric in the quantum region see for instance \cite{Modesto2004,Hossenfelder2010a,Gambini2014a,Olmedo:2017lvt,Malafarina2017,Ashtekar2018,Clements2019,Bodendorfer2019a,Volovik2021,BenAchour2020,Munch2020,Kelly:2020lec}) It was suggested already in \cite{Rovelli2014ps} and has been derived by various quantum gravity research groups, using different methods  \cite{Kelly:2020uwj,Giesel2022,Munch2020,BenAchour2020,Lewandowski2022,Bobula2023,Fazzini2023}, as a credible candidate for the effective metric in the high curvature of a spherically symmetric black hole. For instance, it is uniquely determined by requiring it to satisfy matching conditions with the collapsing matter distribution and keep the Killing symmetry of the Schwarzschild geometry (which in the interior of the horizon is space-like, not time-like as in the exterior.)

This metric has inner and outer Killing horizons. See \cite{Han:2023wxg} for full details. For $m\gg m_{Pl}$, that is $m^2\gg A$, the outer one is at 
\be
r_+ = 2m + O(A/m) \sim r_{Schwarzschild}
\ee
in the classical region. If $m$ is large, this is a negligible modification of the usual Schwarzschild horizon.  While the inner one is at 
\be
r_- = \sqrt[3]{Am/2} + O(A^{2/3}/m^{1/3}) \sim \sqrt[3]{m/m_{Pl}}\,l_{Pl}
\ee
deep inside the quantum region, where the spacetime curvature has Planckian size. These are all also apparent horizons:  they separate trapped, non-trapped and anti-trapped regions \cite{Han:2023wxg}.

\begin{figure}[t]
\begin{center}
\centerline{\includegraphics[width=4cm]{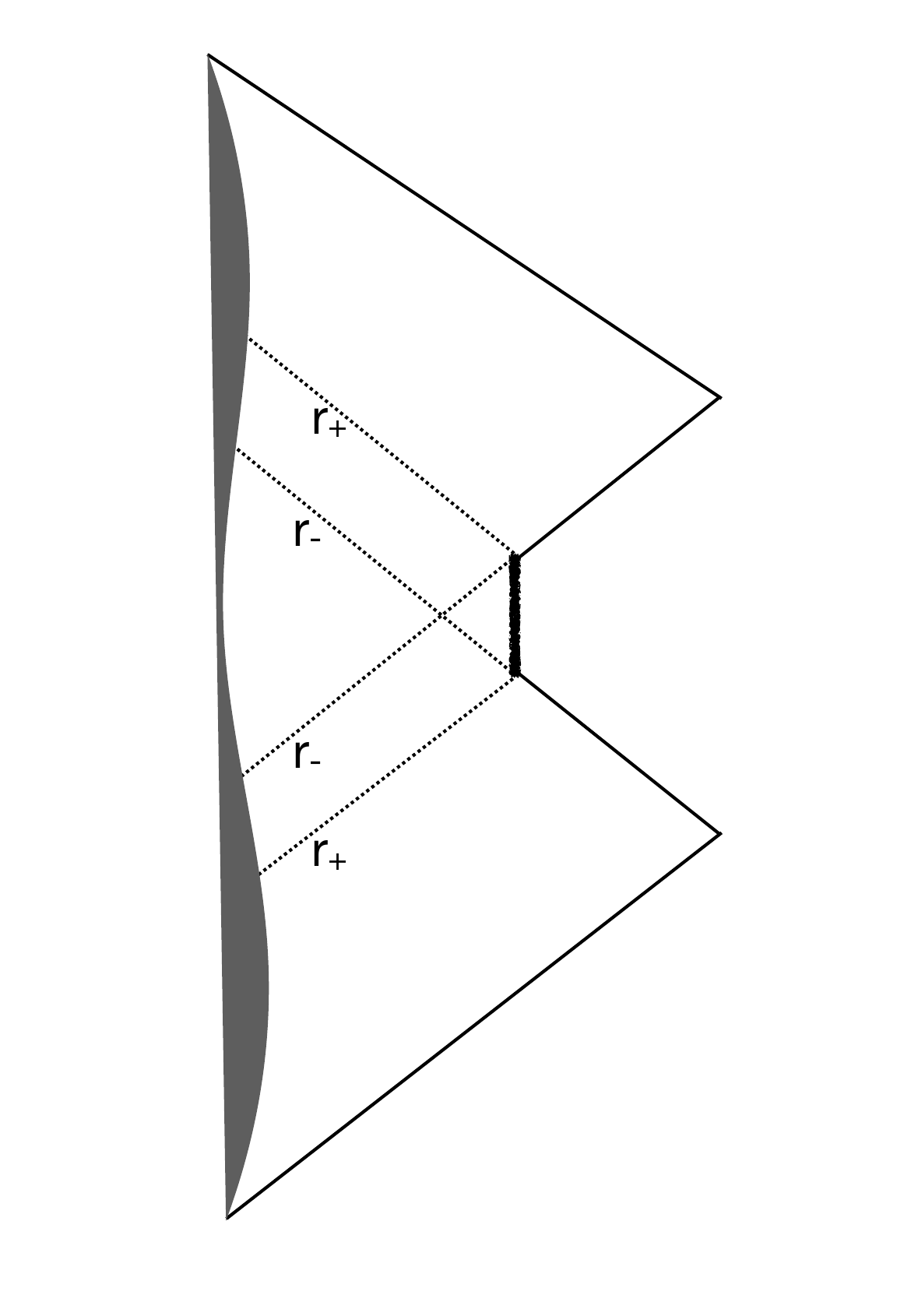}\hspace{.5cm} \raisebox{4mm}{\includegraphics[width=2.7cm]{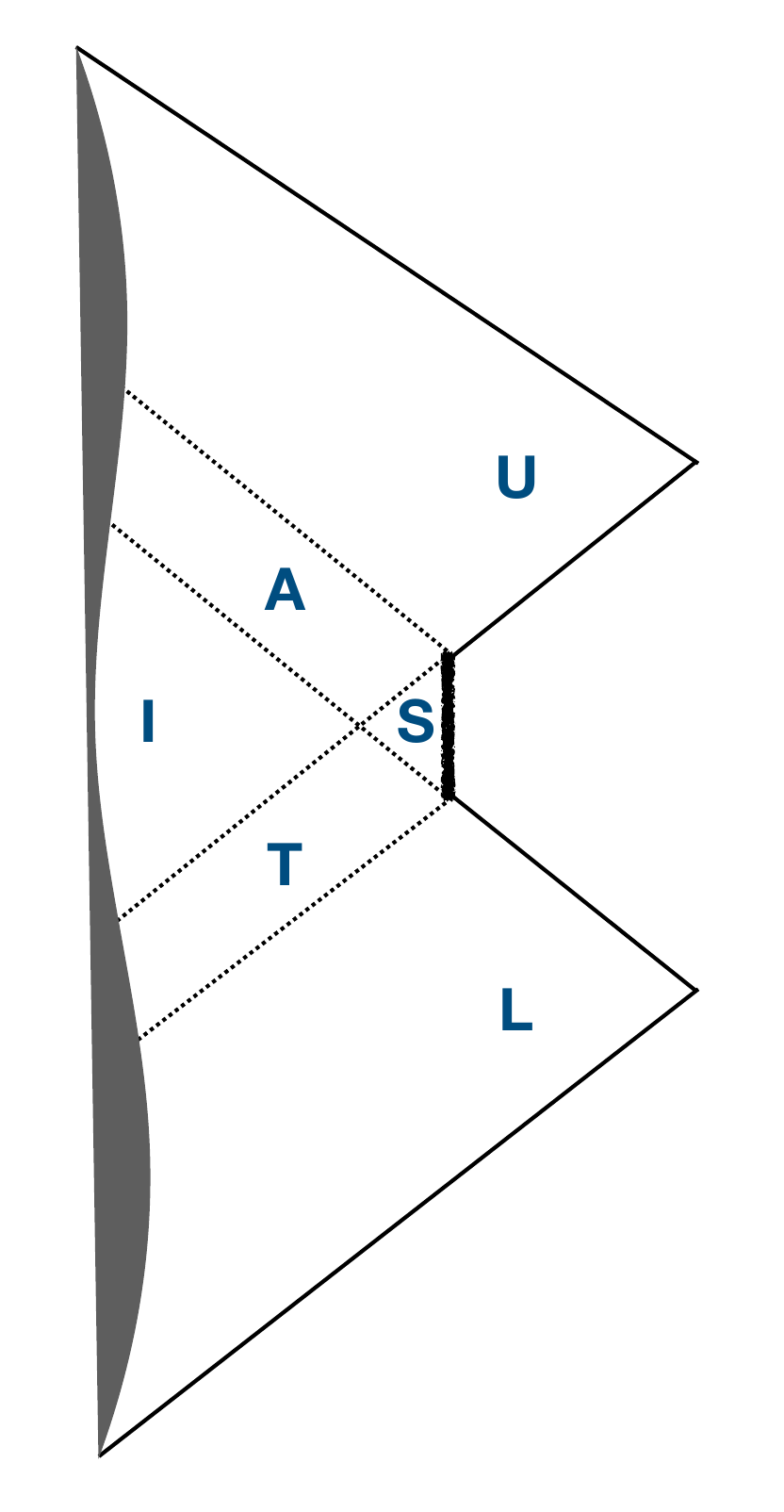}}}
\caption{Conformal diagram of the maximal extension of the spacetime representing the matter distribution and the exterior region defined by eqs.~(\ref{Schwarzschild}) and (\ref{F}).}
\label{CP2}
\end{center}
\end{figure}

The Carter-Penrose diagram of the maximal extension of this metric is depicted in Figure \ref{CP2}, which shows the relative location of these horizons. 

Interestingly this geometry has a global structure (including the inner horizons)  similar to the Kerr, the Reissner-Nordstr\"om, and the general Kerr-Newman black holes. (The full bounce in the Reissner-Nordstr\"om case has been described in \cite{Rignon-Bret2021}, while the Kerr case has not been constructed yet.) 

The maximal extension of this geometry has (i) two asymptotic regions and (ii) a time-like singularity in a high curvature region.  These two features make it an unlikely candidate for describing the actual physics of realistic black holes.     But a surprising result on spherically symmetric solutions of the Einstein's field equations, solves both these difficulties.  This is illustrated in the next section.

\subsection{The exterior metric}\label{exterior}

The Birkhoff's theorem is sometimes  presented as stating that the only spherically symmetric spacetime compatible with the vacuum Einstein's field equations is the Kruskal geometry. At a sufficient distance from a black hole, we expect quantum effects to be negligible and therefore the theorem to hold. Hence we might expect that whatever happens  where quantum effects are negligible must form a subset of a Kruskal geometry.  In the Kruskal metric, there is an anti-trapped region, but it is before, not after the trapped region.  Hence there seems no way to join the metric described above to a single external asymptotic region.  
But the above is wrong.  The reason is that  Birkhoff's theorem is local, not global: it states that a spherically symmetric solution of the vacuum Einstein field equations is \emph{locally}, and not necessarily globally, isometric to the Kruskal metric.  This seems a red herring, but has momentous consequences for understanding the dynamics of quantum black holes.  

\begin{figure}[h]
\includegraphics[height=4.9cm]{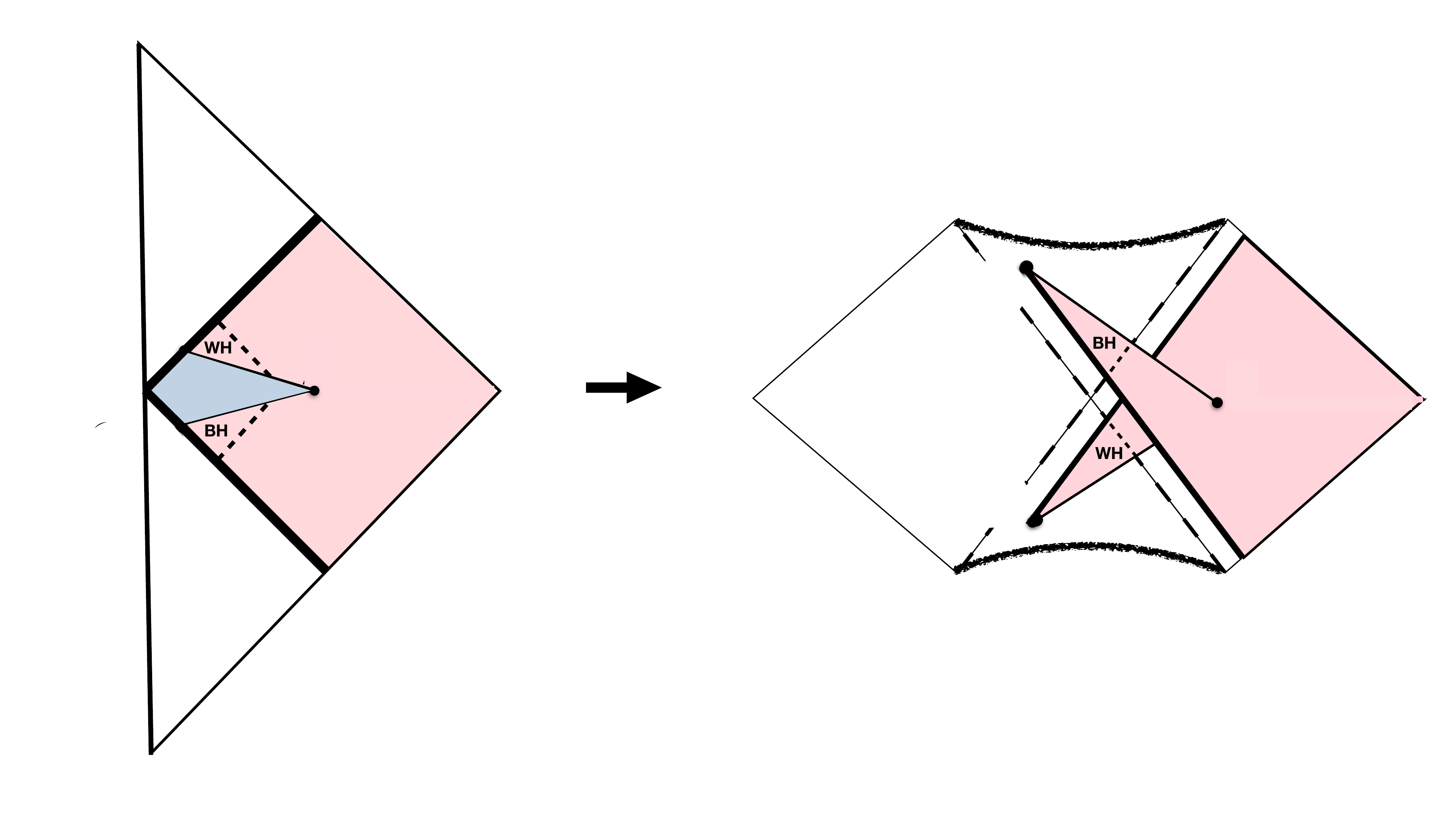}
\caption{The spacetime discovered in \cite{Haggard2014}, describing the collapse of a null shell into a black hole and its bounce out as a white hole (left);  and the way it can be locally isometrically mapped onto the Kruskal spacetime (right). The pink region is an exact solution of the classical Einstein equations.}
\label{foura}
\end{figure}

In fact, as surprisingly realized in \cite{Haggard2014}, there is an \emph{exact} solution of the  Einstein's field equations, with a single asymptotic region, that can surround a quantum transition from a black to a white hole.   The Carter-Penrose diagram of this solution is depicted in the left hand side of Figure \ref{foura}, while the right hand side shows how this can be \emph{locally} isometrically mapped onto the Kruskal spacetime.  The point is that the map is not injective, hence the (pink) classical part of the spacetime is locally isomorphic but not a subset of the Kruskal geometry. 

This solution describes a null shell collapsing into a black hole and bouncing out from a white hole.   The key point is that the black hole (the trapped region) is in the past of the white hole (the anti-trapped region) while in the Kruskal spacetime it is the opposite. This shows that an exact solution of the Einstein equations is compatible with a quantum process happening in a compact high curvature region tunnelling a black into a white hole. 

This same idea solves the two difficulties mentioned at the end of last section, providing a spacetime with a single asymptotic region and no singularities for the bouncing star.   This can be constructed by cutting and pasting the maximal extension of the metric defined by eqs.~(\ref{Schwarzschild}) and (\ref{F}), as follows. 

\begin{figure}[h]
	\includegraphics[width = \columnwidth]{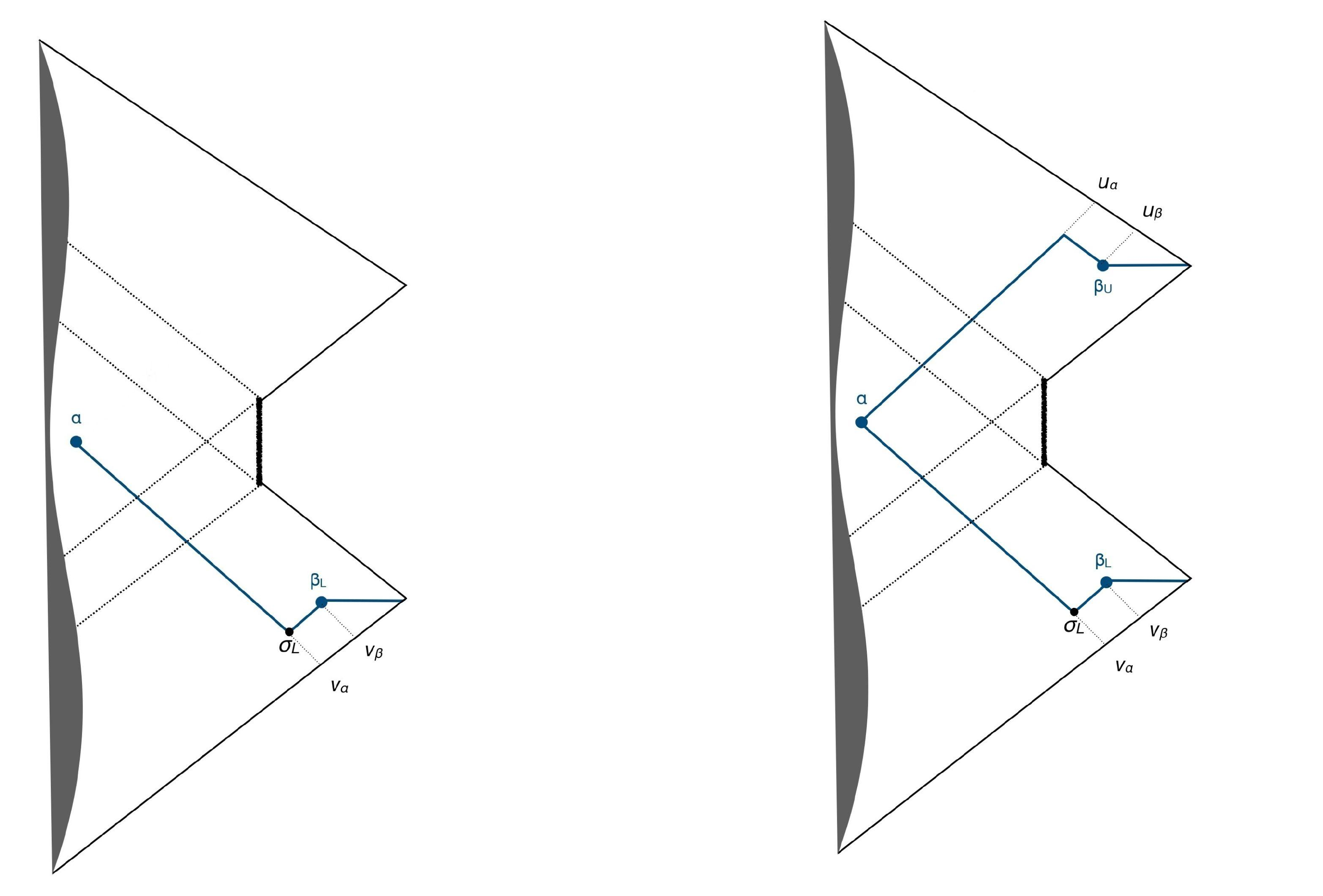}
\caption{The construction of the single asymptotic region spacetime, I: choosing the cut and glue surface}
	\label{cg}
	\end{figure}

First, pick a point $\alpha$ in the interior region, on the surface invariant under time reversal and a point $\beta_L$ outside the horizons, in the first asymptotic region, as in the left panel of Figure \ref{cg}  This choice depends on three parameters: the advanced times $v(\alpha)$ and $v(\beta_L)$ and the Schwarzschild radius of $\beta_L$. Next, consider the blue line of the figure. This has a null portion joining $\alpha$ and $\beta_L$ with their last common past event and a spacelike portion joining $\beta_L$ to spacelike infinity along a constant Schwarzschild time.  Next, draw the line symmetric to this under time reversal, as in the right panel of Figure \ref{cg}.

\begin{figure}[t]
\includegraphics[width = \columnwidth]{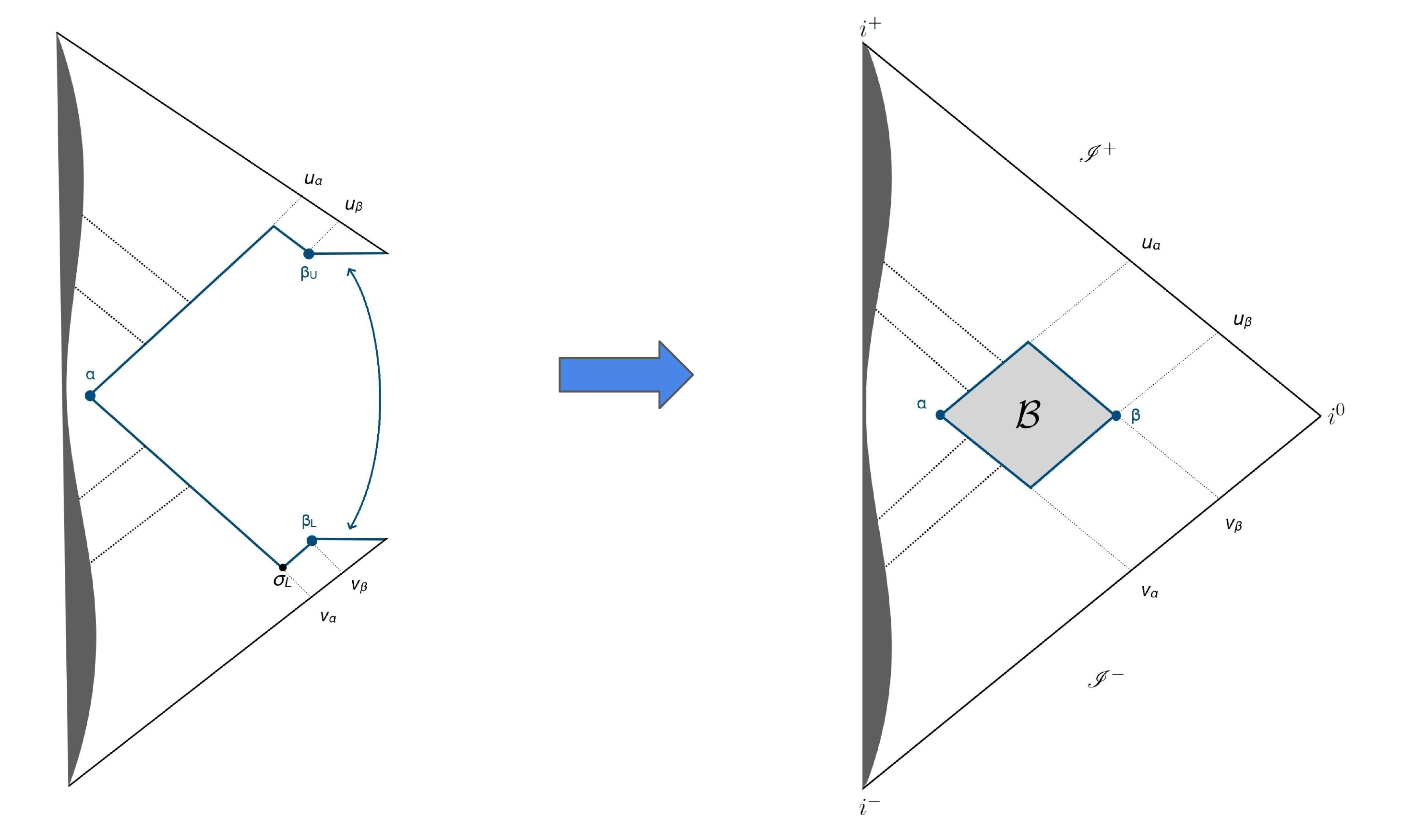}
\caption{The construction of the single asymptotic region spacetime, II: deleting the singular region and gluing the two equal time exterior surfaces.}
	\label{cg2}
	\end{figure}

Next, delete the entire spacetime region enclosed within the blue line, and paste its two spacelike portions as in Figure \ref{cg2}.  This is clearly possible since they are both constant Schwarzschild time surfaces. 

The resulting spacetime has a single asymptotic region, is everywhere locally isomorphic to the maximal extension of the metric (\ref{Schwarzschild}-\ref{F}), and has a hole, depicted as the $B$ diamond of Figure  \ref{cg2}.  In \cite{Han:2023wxg} it is shown that the $B$ region admits a regular Lorentzian metric compatible with rest of the spacetime geometry.  The $B$ region and its physics will be discussed later on. For the moment, let us focus on the rest of the spacetime.

\begin{figure}[b]
    \centering
    \includegraphics[width=7cm]{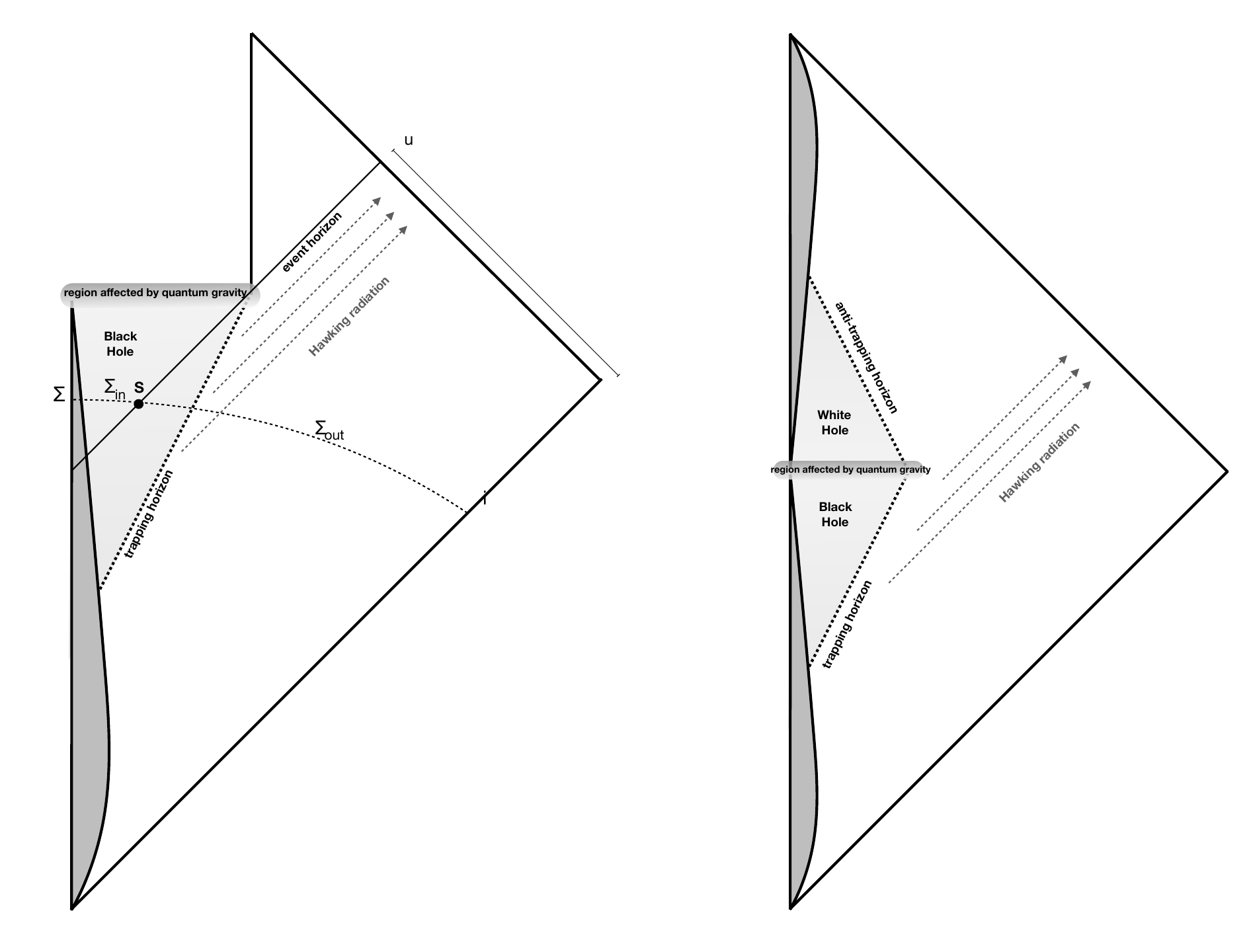}
    \caption{Left: a popular idea of what happens at the end of the evaporation. This is not supported by any theory and contradicts unitarity. On the right, the plausible scenario.}
    \label{b2wSch2}
\end{figure}

The popular idea that at the end of the evaporation a black hole disappears is  not supported by any theory and contradicts unitarity. The above construction offers a far more plausible alternative. See 
 Figure \ref{b2wSch2}.

\begin{figure}[h]
    \centering
    \includegraphics[width=3cm]{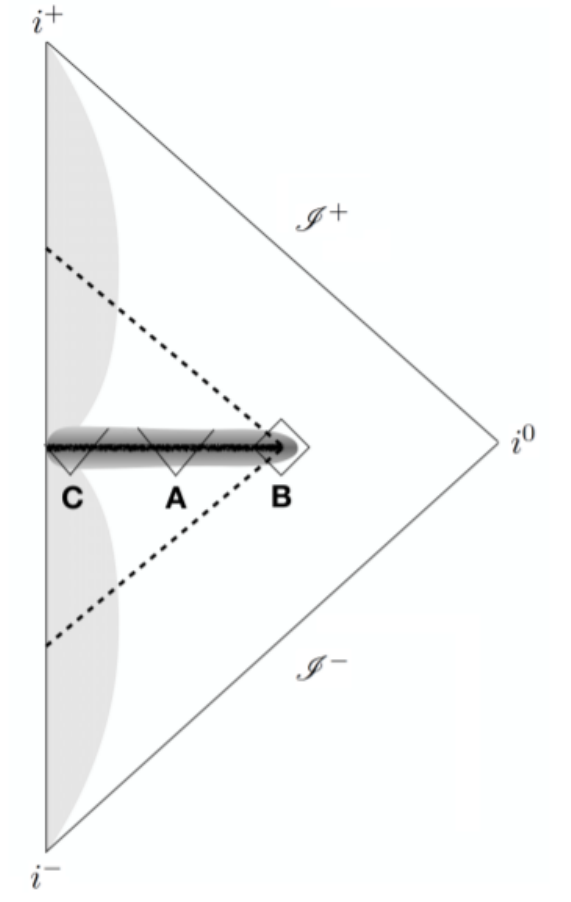}
    \caption{The Carter-Penrose diagram of the black-to-white transition. The dark grey region is the quantum gravity region. The black hole (trapped region) is below the quantum gravity region while the white hole is above. The trapping horizons are the dashed  lines.}
    \label{b2wSch}
\end{figure}

The relation of the resulting geometry with the various quantum gravity phenomena mentioned above is depicted in Figure \ref{b2wSch}.  (See also Figure 8 in \cite{Carballo-Rubio2021}.)

We have constructed the above metric introducing three parameters ($v(\alpha), v(\beta_L))$ and the Schwarzschild radius of $\beta_L$. The resulting geometry (apart from the choice of the metric inside $B$) depends then on these parameters, plus the mass of the star.   Of the three parameters, two simply locate the $B$ region: they determine the minimal and maximal radius of the diamond defining it.   The last one is far more interesting. It determines the global geometry of the spacetime, in the sense in which the radius of a cylinder determines the global geometry of a locally flat cylinder.   Its geometrical and physical interpretation can be seen as follows: it determines the time $T$ at which an observer at large distance from the hole sees the duration of the process.  

\begin{figure}[b]
    \centering
    \includegraphics[width=4cm]{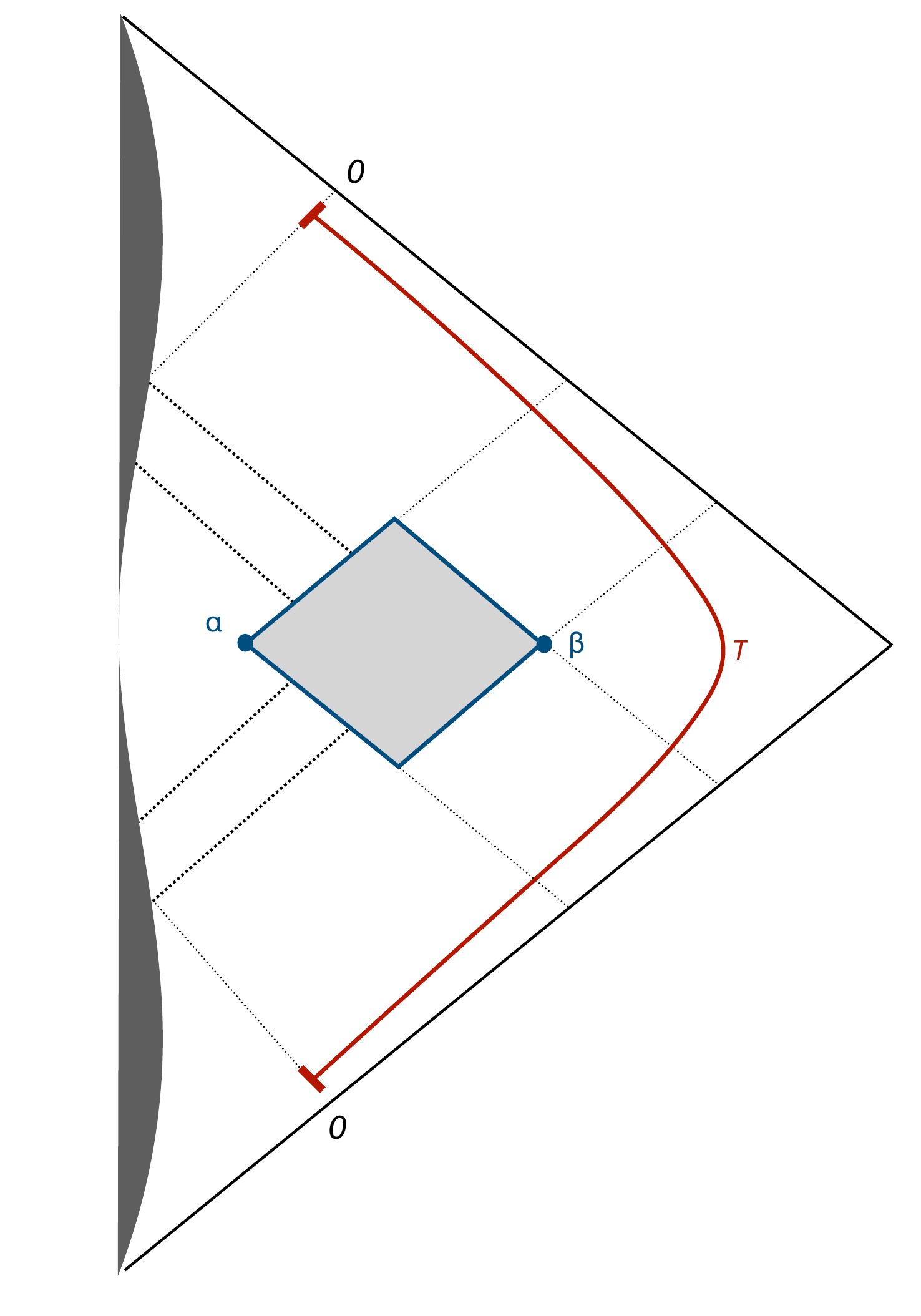}
    \caption{The definition of the duration of the process. The red line represents an observer at fixed radius.}
    \label{time}
\end{figure}
 This is illustrated in Figure \ref{time}, where the red line represents an observer at a fixed radial distance $R$. As shown in \cite{Han:2023wxg}, this duration is 
\be
T=2R+4m\ln{r-2m}-4m\ln\delta.
\ee
The first term on the right-hand side can be interpreted as the back and forth travel time for the light from the observer to the hole.  The second term is the standard relativistic correction.  The last term is independent of the radius and represents a property of the transition itself.   The quantity $\delta$ can be measured by distant observers by measuring $T$ at different values of $R$. A little geometry shows that it  depends on the choice of $\beta_L$ as follows: $r=2m+\delta$ is the radius at which the constant Schwarzschild-time surface passing by $\beta_L$ intersects the surface of the falling star.  This is clearly very close to the collapse (the point where the surface enters entirely in its horizon), if the lifetime of the hole is long.   Thus
\be
\tau=-4m\ln\delta
\ee
can be taken as a definition if the duration of the quantum process, as observed from infinity.   In our construction, it is determined by how the geometry has been constructed by cutting and pasting.  Operationally, it is the observed duration of the process from very large distance.  Physically, it is determined by the quantum theory of the geometry of the $B$ region, to which we now turn.

\subsection{The Boundary Region}

In the previous sections, we have constructed a spacetime geometry that could describe the evolution of a black hole past the region where quantum gravitational effects dominate. How well supported by known physics is this picture? 

The bounce of the matter distribution is supported by the LQG modelling of the quantum dynamics of symmetric spacetimes.   The physical plausibility of the effective metric outside the matter distribution but inside the outer horizons is also supported by various LQG modelling of the quantum dynamics of symmetric spacetimes.  The structure of the $B$ region, on the other hand, has only been guessed as a plausible solution to the requirement of a global dynamics compatible with the external geometry dictated by the classical Einstein equations.   Does it follow from quantum gravity?  Is it allowed by it? 

To address this question we need to step back and think more precisely what we are doing.   First, let us remember once more that for the moment we are disregarding the dissipative effects that break time reversal invariance.  These will be studied in the next section.  Then, consider the fact that we are describing a quantum process. There are different ways of describing quantum processes.   In some regimes, these can be described as corrections to the classical equations of motion. This is not viable for the quantum transition of the horizon, which is a non-perturbative process. One cannot describe quantum tunnelling as a simple $\hbar$ correctios to classical trajectories.  Tunnelling is non-analytical in $\hbar$. 

Another possibility is to describe the evolution of a quantum state. Quantum states that approximate classical configurations, such as coherent states, spread.  We could therefore consider the full quantum state of the geometry after the actual quantum process as a quantum state, namely a quantum superposition of geometries.  This is analogous to describing nuclear radioactivity in terms of a wave function of the emitted particle widely spread in space and time. Correct, but incomplete. What we observe in nuclear radioactivity is not the wave function of the emitted particle,  widely spread in space and time.  Rather, it is a specific space and time location where the particle is detected, say by a Geiger counter.  

In Copenhagen terms, we observe the result of a measurement.  In Many World terms, we observe the position of the particle in one branch, after decoherence.   In Relational Quantum Mechanics, we observe the  metric relative to us, again after decoherence. The theory, of course, does not predict any specific outcome, but it does predict the probability of the  alternatives.  

To compute these, we need to compute transition amplitudes.  This is how we can address the problem of describing the $B$ region in LQG.   

That is, we can compute the transition amplitude for the geometry to make a quantum transition from its value in the past boundary of $B$ to its future boundary (as opposed to not yet making it, as in radioactivity).  Notice that the geometry (both intrinsic and extrinsic) of these two boundaries is known from the construction of the above section.   We have simply to compute the LQG transition amplitude between quantum states approximating these geometries. 

The covariant formulation of LQG is precisely formulated to do this calculation, using approximations provided by  truncation of the degrees of freedom.  The calculation has been performed, under some drastic simplifications by Christodoulou and D'Ambrosio using analytical techniques in \cite{Christodoulou2018,Dambrosio2021,Christodoulou2023}. The result has been confirmed numerically in \cite{Frisoni2023} and \cite{Han2024}. Numerical calculations are been currently developed to go beyond these approximations.  The way these calculations are performed is summarized in the next section.

\subsection{The LQG transition amplitude} % and the Christodoulou-D'Ambrosio result}. 

The covariant formulation of LQG defines transition amplitudes between quantum states of the geometry, in suitable truncations of the numbers of degrees of freedom. The number of degrees of freedom can be truncated by approximating the 3d geometry of the boundary by means of a cellular decomposition.  The two-skeleton of the dual of the decomposition form a graph and the truncated variables can be taken to be $SU(2)$ group elements $U_l$ on the links $l$ of this graph.  The quantum states of the geometry can then be taken to be  functions of the $U_l$'s in the space $L_2[SU(2)^L/SU(2)^N]$, where $L$ and $N$ are respectively the numbers of links and nodes of $\Gamma$, as in lattice Yang Mills theory.  The algebra formed by the group elements $U_l$ themselves and the left-invariant operators is the observable algebra, which has a natural interpretation in terms of functions of the discretized geometry.   A large literature has developed a theory of coherent states for this algebra, representing semi-classical geometrical states. 

Transition amplitudes can be  equally understood as Feynman sums over geometries, or as analogous to lattice Yang-Mills calculations.  Concretely, they are defined in a truncation.  A truncation is here given by a two-complex bounded by the boundary graph.  The amplitude is defined by the product of the vertex amplitudes  
\be
A_v(\psi)=P_{SL(2,C)}Y_\gamma \psi(1\!\!1),
\ee
where $P_{SL(2,C)}$ is the projector on the $SL(2,C)$ invariant part and $Y_\gamma$ is the `simplicity map' from $SU(2)$ to $SL(2,C)$ representations defined, in the canonical basis, by 
\be
|m;j\rangle \mapsto |\gamma j, j; j,m\rangle.
\ee
See \cite{Rovelli2011c,Rovelli2015} for the full details.  This amplitude has been shown to give the Einstein dynamics in suitable limits \cite{Barrett2009} and can be taken as a definition of the covariant LQG theory. 

To apply this general theory to the $B$ region we need to find a suitably simple cellular decomposition of this region.  Steps in this direction have been taken in \cite{Soltani2021}, and a numerical calculation has been achieved in \cite{Han2024}. An  analytical calculation, on the other hand, has been completed only in a simpler setting, where the entire quantum region is discretized.

\begin{figure}[h]
    \centering
    \includegraphics[width=3cm]{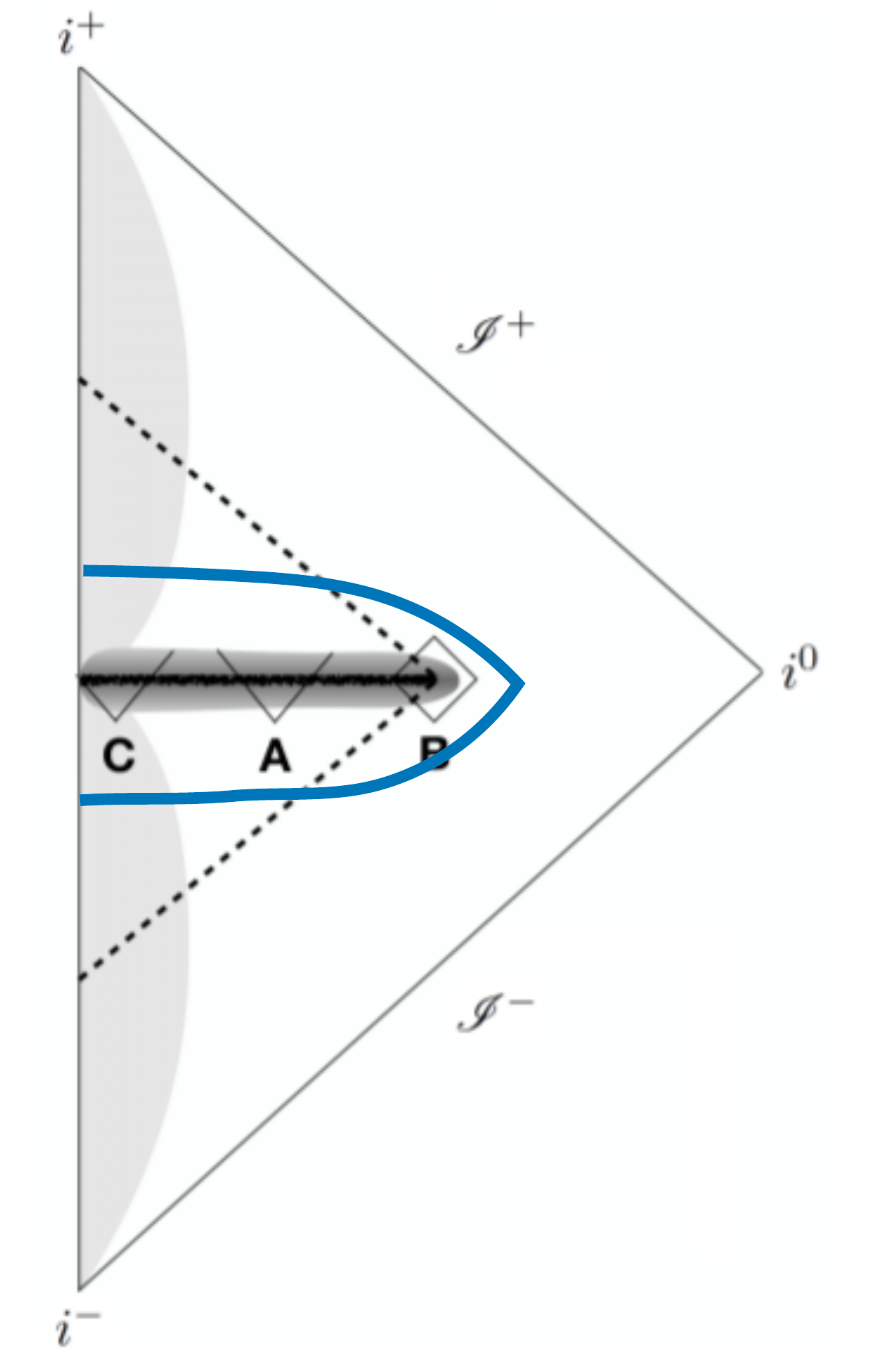}
    \caption{The boundary of the quantum region for the simplified calculation of the transition amplitude.}
    \label{b1}
\end{figure}

The complexity of the calculation on the $B$ region alone is given by the topology of $B$, which is the product of a two-sphere and a disk.  The disk is the product of a finite time interval and a finite radial interval and it is delimited by an exterior two-sphere $S_+$ that surrounds the horizon and by an interior two-sphere $S_-$ surrounded by the horizon (sitting on the bounce radius of the transition of the internal geometry of the black hole). The two two-spheres $S_+$ and $S_-$ split the boundary $\Sigma$ of $B$ into a past component $\Sigma^p$ and a future component $\Sigma^f$.  

\begin{figure}[h]
    \centering
    \includegraphics[width=7cm]{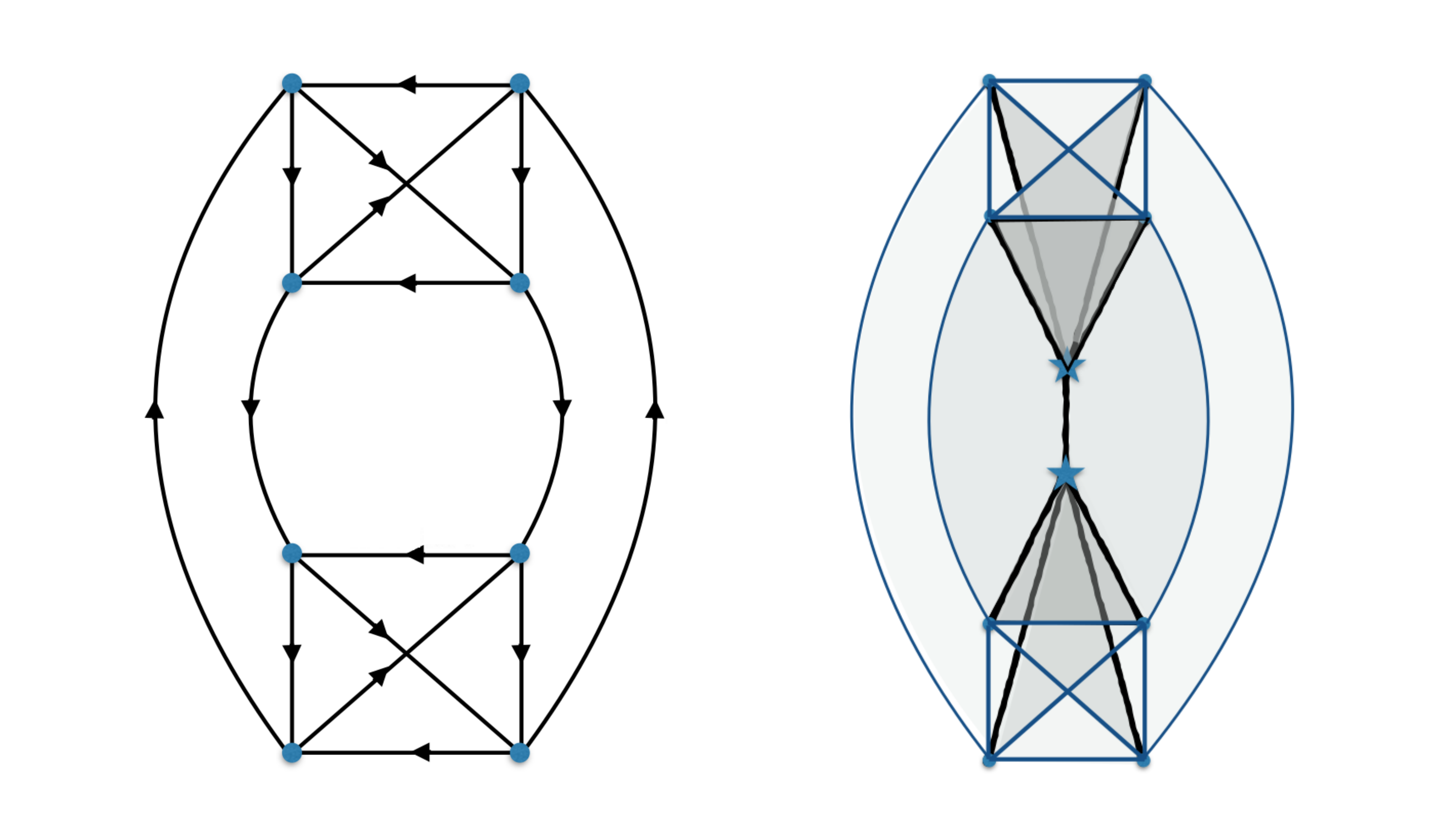}
    \caption{The graph of the spin network and the two-complex of the spinfoam for the calculation of the transition.}
    \label{spf}
\end{figure}

Considering instead the entire quantum region simplifies the topology. The boundary of the quantum region can be take to be as depicted in blue Figure \ref{b1}. The tip, at the largest radius, is a two-sphere, chosen  outside the Planckian curvature region.  The lowest (past) part of the boundary and the upper (future) one are both 3d balls.  Each of these can be triangulated with four tetrahedra and the interior can be triangulated with two four-simplices sharing an internal tetrahedron, as depicted in Figure  \ref{spf}.

We refer to  \cite{Christodoulou2018,Dambrosio2021,Christodoulou2023} for the analytical calculation.  

The numerical  calculations have been performed using  different techniques. \cite{Frisoni2023} has analyzed the simpler geometry, utilizing the sl2cfoam-next code for computing spinfoam transition amplitudes. A good introduction to this, and full references, is in \cite{Dona2022a}.  The code can explore the low-spin (small-mass) regime of the amplitude. 
On the other hand, \cite{Han2024} utilizes the complex saddle point method for computing spinfoam amplitudes, which can explore the high-spin (large-mass) regime \cite{Han2020,Han2020,Han2023,Han2024a}.   A common result of all these  investigations (that contradicts some previous expectations) is that the transition amplitude gives a transition probability proportional to 
\be
P\sim  e^{-\frac{Gm^2}{c\hbar}} \sim e^{-(\frac{m}{m_P})^2}.
\ee
Notice that this is indeed non-analytic in the Planck constant. Therefore it is a genuine non-perturbative result that cannot be obtained in conventional perturbation theory expanding around a classical solution. The exponential is characteristic of quantum tunnelling phenomena. The transition, in fact, is forbidden classically (classically the geometry evolves into a singularity) and is a characteristic quantum tunnelling effect. 

Another way of understanding this result is to connect to the asymptotic behaviour of the amplitudes 
\be
W\sim  e^{iS_{Regge}}\sim  e^{i\sum_fj_f\Theta_f(j)},
\ee
where $S_{Regge}$ is the Regge action: a sum over the faces of the face spin times the dihedral angle of the face, which is a function of all the spins, determined by the interpolating flat geometry. Here there is no interpolating geometry and the angles turn out to be imaginary, giving
\be
W\sim   e^{-c\sum_f j_f}\sim   e^{-c\sum_f A_f},\sim   e^{-c\; m^2}.
\ee
The simplest way to interpret this result is in analogy to nuclear radioactivity: as a transition probability per unit of (here Planck-) time.  The immediate consequence of this result is that the transition probability is exponentially suppressed if the mass of the hole is larger than the Planck mass.  

Anticipating the discussion about dissipative effects, we can see that toward the end of the Hawking radiation, $m$ decreases and approaches $m_P$. At this stage the transition becomes increasingly probable, going to probability unity when the mass becomes actually Planckian. 

In other words, the calculation indicates that the transition can happen and it is likely to happen, but only at the end of the Hawking evaporation,  giving birth to a white hole with near Planckian mass.

\subsection{White holes} \label{WhiteHoles}

Before addressing the dissipative aspects of the life of a hole, let us pause to clarify a few properties of the white holes.  White holes, like black holes, are exact solutions of the Einstein's equation.    Precisely like black holes, they have long been expected  to not play any role in our universe, by a majority of physicists. The situation has changed for black holes in the last decades. It has not yet equally changed for white holes, but it might soon. 

\begin{figure}[h]
    \centering
    \includegraphics[width=7cm]{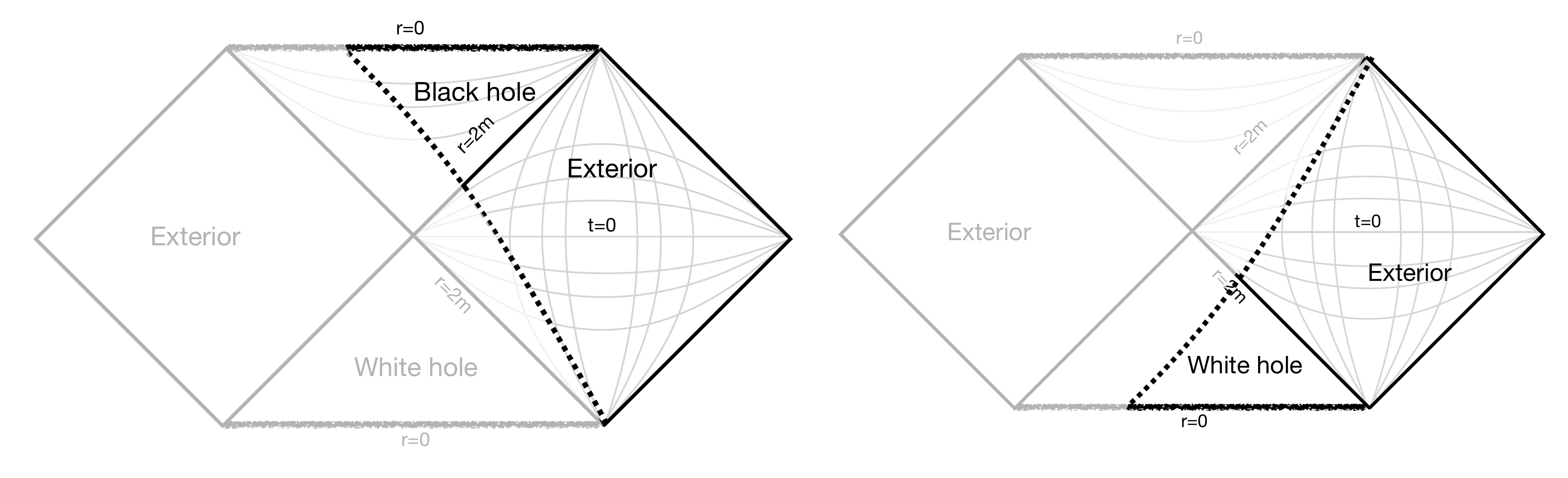}
    \caption{The spacetimes of a black and a white hole (outside the respective stars, as subsets of the Kruskal geometry).}
    \label{wb}
\end{figure}

A white hole spacetime is simply the time reversal of a black hole spacetime.  For instance, a classical black hole formed by a collapsing matter distribution  and its time reversal, a white hole from which matter distribution emerges, are depicted in Figure \ref{wb}.

The reason black holes have been traditionally taken more seriously as candidates for real objects than white holes is that we have a well understood classical scenario for how a black hole can form, but not so for a white hole.   Since the two are the time reversal of one another, this implies that we have a well understood classical scenario for how a white hole can end, but not so for a black hole.   The end of a black hole is definitely a quantum phenomenon, because it involves high curvature regions, where quantum theory cannot be neglected.  The same must be true for the birth of a white hole.   

All this points to a very simple possibility: the end of a black hole is the birth of a white hole, via a quantum transition process.   This is precisely the scenario we are studying here.   A white hole can be originated by a dying black hole. 

The difference between a black hole and a white hole is not very pronounced.  In fact, observed from the outside (say from the exterior of a sphere of radius $r=2m+\epsilon>2m$, where $m$ is the mass of the hole) and for a finite amount of time, a white hole cannot be distinguished from a black hole.  

This is clear from the usual Schwarzschild line element, which is symmetric under time reversal, and therefore describes equally well the exterior of a black hole and the exterior of a white hole.  Equivalently, zone II of the maximal extension of the Schwarzschild solution is equally the outside of a black hole and the outside of a white hole (see Fig. 1, Left). Analogous considerations hold for the Kerr solution. The continuation of the external metric of a stationary Kerr or Schwarzschild spacetime inside the radius $r=2m+\epsilon$ contains both a trapped region (a black hole) and an anti-trapped region (a white hole).

\begin{figure}[h]
	\includegraphics[width = .33 \columnwidth]{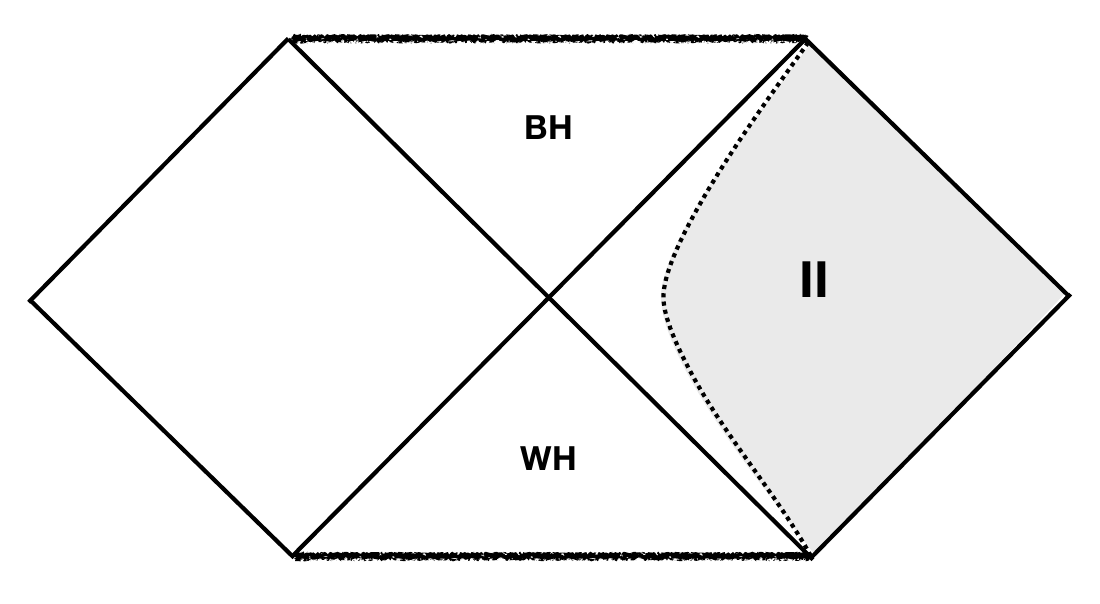}\hfil
	\includegraphics[width = .33 \columnwidth]{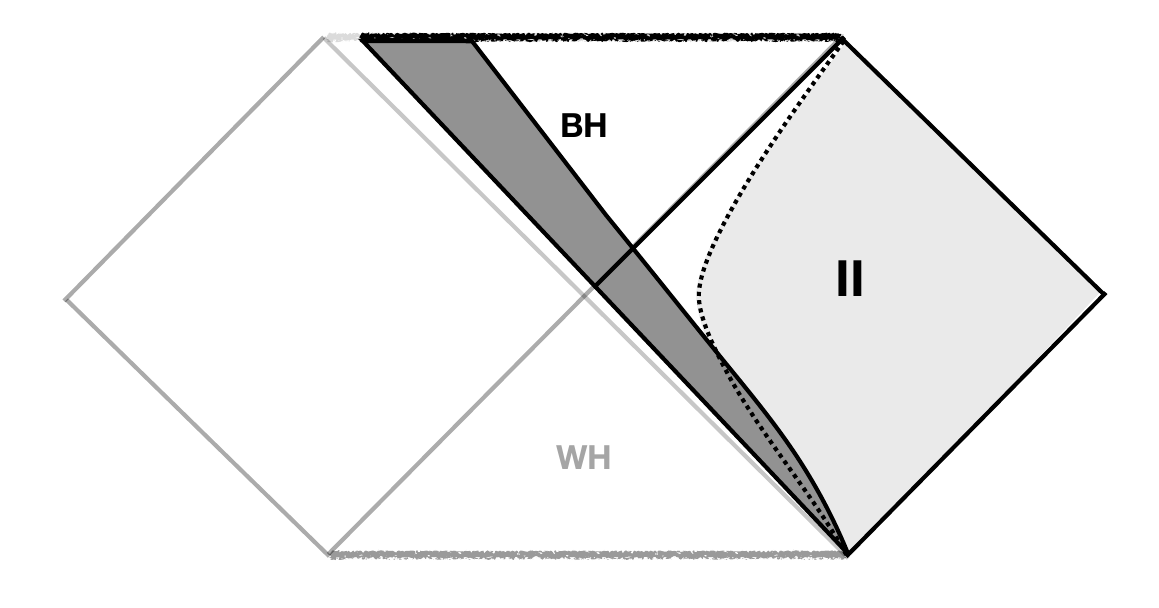}\hfil
	\includegraphics[width = .33 \columnwidth]{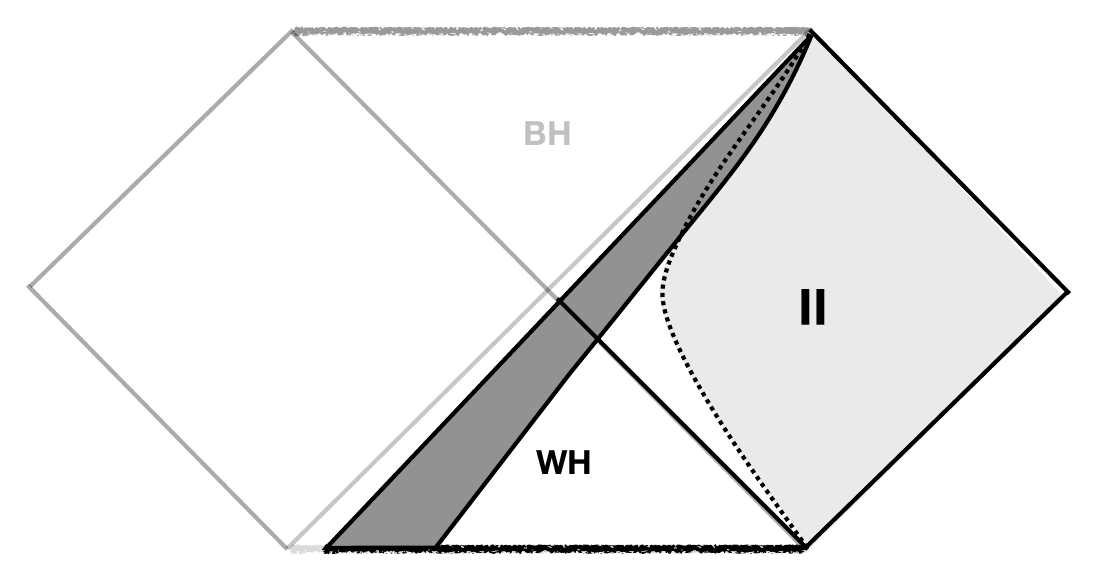}
\caption{Left: in the extended Schwarzschild spacetime, which is stationary, the (light grey) region outside $r=2m+\epsilon$ (dotted line) is equally the outside of a black and a white hole. Center: A collapsing matter distribution (dark grey) replaces the white hole region (``WH") in the non-stationary collapse metric.  Right: The time-revered process. The difference between the last two can only be detected looking at the past, or the future.}
	\label{Kruskal}
	\end{figure}
	
What distinguishes then the objects we call `black holes' from `white holes'? The objects in the sky we call `black holes' are described by a stationary  metric only approximately and for a limited time.   We expect that (at least) in the past their metric was definitely non-stationary, and they were produced by gravitational collapse.  The energy contained inside $r=2m+\epsilon$ was less than $m$ in the past and the continuation of the metric inside this radius contains the trapped region, but not the anti-trapped region, which is instead replaced by the region describing the collapsing star.  Therefore seen from the outside a `black hole' (as opposite to a `white hole') is only characterized by the fact that in the past it does not have an anti-trapped region (see Fig. 1, Center). Vice versa, a white hole is an object that from the exterior and for a finite time is indistinguishable from a black hole, but in the future ceases to be stationary, the amount of energy inside $r=2m+\epsilon$ decreases, and there is no trapped region in the future (see Fig.1, Right).  All this regards classical gravity only.

\section{Dissipative aspects of the transition}  \label{II}

\subsection{Black hole lifetime}

The most relevant  dissipative phenomenon in the life of a black hole is the Hawking radiation. This is a markedly irreversible process that dissipates the energy of the collapsed matter distribution into the heat of the emitted thermal radiation. 

The lifetime $\tau_{BH}$ of a black hole is known by Hawking radiation theory. It can be estimated as follows.  The Hawking radiation is thermal, namely it has a Planck spectrum.  The peak of the spectrum is on a wavelength $\lambda$ that is determined by the only length parameter in the problem: the Schwarzschild radius $2Gm/c^2$.  Therefore the radiation is mostly composed by quanta of frequency 
\be
\nu=c/\lambda\sim \frac{c^3}{Gm}
\ee
These have an energy $E=h\nu$ and therefore (being thermal) a temperature $T$ given by
\be
kT=E\sim\frac{c^3\hbar}{Gm}
\ee 
This is (up to a numerical factor) Hawking's temperature. The emission of the horizon can be modelled  as the emission of a sphere at this temperature with the area of the horizon. This gives an emitted power (going back to natural units)
\be
P=\frac{dm}{dt}\sim AT\sim m^2 m^{-4}=m^{-2}.
\ee
This is a differential equation for $m(t)$ that can be immediately integrated giving $m^3\sim t$. Therefore we expect that the finite lifetime of a black hole due to the Hawking evaporation is of the order
\begin{equation}
\tau_{BH}\sim m_o^3
\end{equation}
where $m_o$ is the \emph{initial} mass of the hole.\footnote{The effect of the back reaction on the metric of the full bounce has been studied in \cite{martin2019evaporating}. See references there.}

The evaporation shrinks the horizon, bringing it close to Planckian size, where the black-to-white transition has high probability to happen.  The white hole generated by the process then has a horizon of Planckian size.  (This lifetime can be shorter if quantum gravity fluctuations trigger an earlier tunnelling. This could be as early as $\tau_{BH}\sim m_o^2$ as suggested in \cite{Haggard2014,Haggard20162}.  
%(see also \cite{Gregory1993,Casadio2000,Casadio2001,Emparan2003,Kol2004}). 
 Here we only focus on the possibility that the transition happens at the end of the evaporation.)

However, the fact that the black hole has evaporated does not mean that it is ``small" in any sense of the word.  In fact, its interior is vast.   To understand this point, a detour on the size of the interior of black holes is important, as this is a frequently misunderstood and misleading issue. 

\subsection{How big is the interior of a black hole?}

In Minkowski spacetime, we say that a space-like 2-sphere $S$ of radius $r$ encloses a space-like ball of volume $\frac43\pi r^3$.   As a 3d surface in Minkowski space, this ball is characterized by two properties: it is linear, and it is the surface that maximizes the volume among those bounded by $S$.   Any deformation, in fact, is time-like and \emph{decreases} the volume.  

In a curved Lorentzian manifold,  linearity is meaningless, but we can still talk about the volume enclosed in a spacelike (topological) 2-sphere $S$ by defining it as the volume of the 3d surface that \emph{maximizes} the volume, among all 3d surfaces bounded by $S$.   This definition, introduced in \cite{Christodoulou2015}, allows us to talk of the interior volume of a black hole in a rigorous manner, and provides a simple intuition of the internal geometry of a black hole by defining a natural foliation of this geometry.

Consider therefore the spherical symmetric spacetime of a collapsing matter distribution and choose a retarded time $v$.  The intersection of the retarded time with the horizon defines a 2-sphere $S_v$.  Let $V(v)$ be the volume of the maximal-volume spacelike ball bounded by $S_v$.   This is the volume of the interior of the black hole at the retarded time $v$ on its horizon.  The dependence of this volume on $v$ has been computed both for an eternal horizon and for an evaporating one \cite{Christodoulou2015,DeLorenzo2016,Bengtsson2015,Ong2015,Wang2017,Christodoulou2016a}:  it grows linearly in $v$. 

Specifically \cite{Christodoulou2015,Christodoulou2016a} that, at advanced time $v$ after the collapse, a black hole with mass $m$ (disregarding evaporation) has interior volume 
\be
V\sim 3\sqrt{3} \pi\; m^2 v
\ee
for $v\gg m$.

Recall that the interior of the horizon is never stationary: it is dynamical. (The common expression ``stationary black hole" means a black hole whose \emph{external} geometry is stationary.)  In the natural foliation considered above, the internal volume of the black hole increases steadily in time.  More specifically, the interior is like a tube whose radial dimensions shrink in time while its longitudinal dimension increases, in such a way that the total volume increases. 

This implies that  an old evaporated black hole has a small horizon \emph{but a huge internal volume}.   Ignoring this fact has nourished a wrong intuition about the geometry of evaporated black holes. Contrary to what is sometimes believed, these are genuinely different from young black holes with the same mass.

In other words, we must not be fooled by the no-hair theorem. This theorem states that two black holes that have the same mass, charge and angular momentum, classically settle on the same \emph{external} geometry, but certainly does not say that they have the same interior! Older ones are bigger.   In fact they can have vastly different interiors. Accordingly, in the quantum theory, the full quantum state of a black hole will certainly not be determined by external quantities like mass, charges and angular momentum only: there will be quantum numbers $v_i$ distinguishing holes with the same exterior but (possibly vastly) different interiors.  

\begin{figure}[h]
\includegraphics[width = .90 \columnwidth]{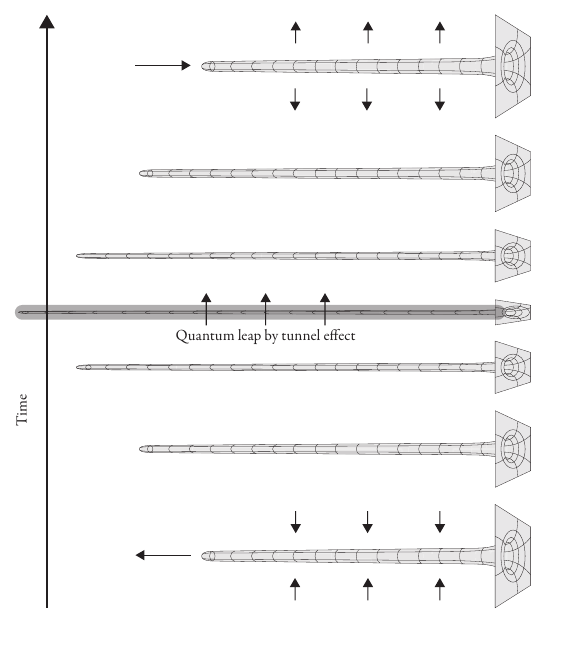}
\caption{A sketch of evolution of the internal geometry of the hole at the bounce (up to some quantum gravity phenomena, see later).}
	\label{geometry}
	\end{figure}

At the moment of the quantum transition therefore an old (previously) macroscopic black hole is an  extremely long geometrical entity.  It is radially shrinking, but growing in length.  The transition is a bounce that reverses the process: the interior of a white hole is expanding radially, and shortening. See Figure \ref{geometry}.

\subsection{Remnants and their lifetime}

The argument of the previous section suggests that at the end of the evaporation a black hole  undergoes a quantum transition to a white hole with a Planckian-size horizon and a vast interior.  

The possibility of remnants with a similar structure was considered in the 1990's \cite{Giddings1992b}. What was not realized at that time is that classical GR does in fact predict the existence of objects with these properties: white holes; and quantum theory could account for their formation at the end of the evaporation and (as we shall see below) for their stability.   

Before studying  in more detail the quantum structure of these remnants, let us ask what could be their lifetime.

A number of arguments  suggest that this should be long. First, the  Einstein equations state that it takes a  long (advanced) time to generate a black hole with a  large interior. Time reversing, they also state that it takes a  long (retarded) time to dissipate a white hole with a large interior.

The second argument concerns the information trapped inside the hole. The Hawking radiation is genuinely thermal: it is describe by a quantum state which is not pure: it is a density matrix. This is because the Hawking radiation that escapes to infinity forms together with negative radiation that falls inside the hole.  The two are entangled, therefore the part escaping to infinity has von Newman entropy. The total entropy of the Hawking radiation is of the order of the Bekenstein-Hawking entropy
\be
S_{BG}=\frac{A}{4}.
\label{Sbg}
\ee  
where $A\sim m_o^2$ is the \emph{initial} area of the black hole, before the beginning of the evaporation.  Some scientists expect that the total entropy of the Hawking radiation starts decreasing at Page time because late Hawking quanta are correlated with early ones.   We are not convinced by this expectation for reasons that we explain in detail later on.  
Hence an amount of information of the order $S_{BH}$ is trapped inside the hole during the quantum transition.  This must be emitted later in the form or radiation by the remnant. The remnant has total energy of the order of the Planck mass, namely of order unity in natural units. It must therefore be able to emit a large amount of information by emitting very small energy. 

To emit a lot of information within small energy, we need a large number of very low-energy quanta. These must have very low frequency, and hence require a long time.   As shown in \cite{Preskill1993,Marolf2017a,Kazemian2022a} (and with an argument detailed in Section \ref{sec:emission}) this leads to a lower bound in emission time: 
\begin{equation}
\tau_{W\!H}\sim m_o^4.
\end{equation}
Here $m_o$ is the mass of the hole before the evaporation (which determines the amount of information in the hole).  

Notice that the state of the black hole at some given (advanced) time, which determines its full future evolution, is not specified only by its current mass $m$, but also by the internal geometry, which in turn is determined by the initial mass of the black hole. 

We can account for this by writing the quantum state of the black hole at some given time in the form $|m_o,m\rangle_B$, where the first quantum number is the initial mass, which determines the size of the interior, while the second is the current mass which determines the area $A=16\pi m^2$ of the horizon on the given time slice and decreases in the evaporation. We do not need to keep track of other possible quantum numbers here.   

At formation, the hole is in the state $|m_o,m_o\rangle_B$, then $m$ decreases by Hawking evaporation until the state $|m_o, m_{P\ell}\rangle_B$.  This states tunnels to a white hole state with the same quantum numbers, which we denote $|m_o, m_{P\ell}\rangle_W$. The tunnelling process itself from black to white is short and takes a time of the order of the current mass \cite{Christodoulou2018,Barcelo2016}.  Here is therefore a first account (which we correct below) of the full life cycle of a gravitationaly collapsed object
\onecolumngrid
\begin{eqnarray}
\xrightarrow[\text{collapse}]{} |m_o,m_o\rangle_B  \xrightarrow[\text{black hole}]{\tau_{W\!H}\sim m_o^3}
 |m_o,m_{P\ell}\rangle_B  \xrightarrow[\text{tunnelling}]{\tau_{T}\sim m_{P\ell}}
% \hspace{1cm}
%  \nonumber\\\nonumber 
% \hspace{1cm}
  |m_o,m_{P\ell}\rangle_W \xrightarrow[\text{white hole}]{\tau_{W\!H}\sim m_o^4} 
|m_{P\ell},m_{P\ell}\rangle_W  \xrightarrow[\text{full dissipation}]{ }. 
\end{eqnarray}
\twocolumngrid

These are time scales for a distant observer.   If $m_o$ is large, the process is very long in the time of a distant observer.  But it is extremely short (of order $m_o$, which is the time light takes to cross the radius of the star) if measured on the bouncing matter distribution itself. The huge difference is due to the extreme gravitational time dilation \cite{Haggard2014}. Time slows down near high density mass. An observer (if capable of resisting the tidal forces) landing on a Planck matter distribution will find herself nearly immediately in the distant future, at the time where the black hole ends its evaporation. The \emph{proper} lifetime of a Planck matter distribution is short: from its own perspective, the matter distribution is essentially a bounce.  A black hole is a shortcut to the distant future. 

This is a compelling scenario, but it still needs to be corrected. 

\subsection{Instability}

A classic macroscopic white hole is an unstable solution of the Einstein equations (see Chapter 15 in \cite{Frolov2012} and references therein).   This means that there are solutions with initial data that are arbitrarily close to the white hole initial data, but have a qualitatively different future.  

The qualitative different future is the formation of a black hole in the future of the white hole.  That is, a white hole is unstable toward becoming a black hole.  The transformation of a white hole into a black hole is similar to the process described by the vacuum Kruskal spacetime.  

To see why there is this instability it is easier to address the time reversed scenario first, because we have a better intuition about it: to show that solutions with \emph{final} data arbitrarily close to those of a \emph{black} hole, must have a qualitatively different past from the black hole. Or in other words, that the data on future null infinity on a black hole spacetime cannot admit certain arbitrary small variations.

To see this, consider a black hole of mass $m$, formed by a collapsed star.  In the unperturbed solution, no energy reaches future null infinity (here we are considering classical physics: no Hawking radiation).  Now, consider a perturbation of this spacetime in which there is a small spherically symmetric pulse of null radiation with total energy $\epsilon$ arriving at null infinity.   Here $\epsilon$ can be arbitrarily small.  In order to arrive at null infinity, this radiation must be emitted by the surface of the matter distribution before this enters the horizon.  Say it was emitted when the radius of the matter distribution was $2m+\delta$. But this  must also be outside the Schwarzschild radius of the matter distribution plus the energy to be emitted, otherwise it could not reach null infinity either.  Hence necessarily $2m+\delta>2m+\epsilon$ that is, we must have $\delta>\epsilon$. Now let $u_\delta$ be the retarded time of the emission point. Necessarily, the energy pulse of energy $\epsilon$ must reach null infinity \emph{before} $u_\delta$. Therefore, no perturbation with energy $\epsilon$ can reach null infinity \emph{after} $u_\delta$. That is, we cannot perturb the final data arbitrarily. However small is $\epsilon$ there are locations where it cannot get to.  

Can a perturbation with energy $\epsilon$ reach null infinity \emph{after} $u_\delta$, with some different past?   Yes it does! A white hole of mass $2m+\epsilon$ can emit the pulse and then become a black hole of mass $m$! In this case, the pulse is never at a radius where it can trigger an increase of the size of the blackhole. 

Time reversing this scenario shows immediately why a classical eternal white hole of mass $m$ is unstable: any arbitrary small pulse originating from past null infinity sufficiently early in time will bring enough energy to be inside the Schwarzschild radius before reaching the matter distribution emerging from the white hole, thus triggering the formation of a black hole. 

In other words, the spacetime depicted in the Center panel of Figure \ref{Kruskal} does not change much under a small arbitrary modification of its initial conditions on past null infinity; but it is drastically modified if we modify its final conditions on future null infinity.  This is intuitively simple to grasp: if we sit on future null infinity and look back towards the hole, we see a black disk. This is the final condition. A slightly perturbed final condition includes the possibility of seeing radiation arriving from this disk. This is impossible in the spacetime of the Center panel of Figure \ref{Kruskal}, because of the huge red shift of the radiation moving next to the horizon, but it is possible in the Left spacetime, because the radiation may have crossed over from the other asymptotic region.  

The same is true for a white hole, reversing the time direction.  In the spacetime depicted in the Right panel, with some radiation, there is necessarily a dark spot in the \emph{incoming} radiation from past null infinity.  If we perturb this configuration, and add some incoming radiation to this dark spot, the evolution generically gives the spacetime of the Left panel.  

Physically, what happens is that this radiation moves along the horizon, is blue shifted, can meet radiation coming out of the white hole and this is more mass than $m$ at a radius $r\sim 2m$: it is mass inside its Schwarzschild radius.  At this point the region is trapped, and a black hole forms.  Consequently the evolution of the perturbed initial conditions yields the spacetime of the Left, not the one of the Right: the white hole is unstable and decays into a black hole.  This is the standard ``instability of white holes" in classical GR.

\subsection{Planckian Remnants}  \label{Remnants}

How does the instability discussed in the last paragraph affect the remnants formed at the end of a black hole evaporation? 

As observed in \cite{Bianchi2018e}, the wavelength of the perturbation needed to trigger the instability must be smaller than the size of the hole.  To make a Planck size black hole unstable, therefore, we would need trans-Planckian radiation, and this is likely not be allowed by quantum gravity.  

But let's nevertheless assume that a Planck-scale white hole is unstable. Its decay mode, as explained above, is into a Planck-scale black hole. Using the notation introduced above, this is the process
\begin{equation}
 |m_o,m\rangle_W  \xrightarrow[\text{instability}]{}  |m_o,m\rangle_B. \label{inst}
 \end{equation}
The quantum gravity process discussed earlier in the paper is 
\begin{equation}
 |m_o,m\rangle_B  \xrightarrow[\text{tunnelling}]{ }  |m_o,m\rangle_W . \label{tunn}
\end{equation} 
In quantum mechanics,  
two such processes imply that a system free to lose energy into radiation will settle on its lowest available energy eigen-state, which will be a superposition of the two states, of the form 
 \begin{equation}
 |m_o,m\rangle =\alpha|m_o,m\rangle_W  + \beta   |m_o,m\rangle_B. 
\end{equation}
That is, the remnant will settle into a superposition of black and white hole.\footnote{The possibility of oscillation between black and white hole states has been also considered in \cite{Barcelo:2015uff,Barcelo2016,Garay2017}. In these works, the tunnelling probability when $m\gg m_{P\ell}$  was estimated to be large on the basis of a black-hole lifetime computation giving $\tau_{BH}\sim m_o$. In \cite{Christodoulou2018}, this computation has been rather reinterpreted as giving the duration of the tunnelling transition itself.}

Since the black and white holes are indistinguishable from the exterior, this superposition has no effect on the exterior. It is only a quantum spread of the geometry on and just around the horizon. In the next section we describe it more precisely.  

Minimal energy, for the horizon, means  minimal area.  However, vanishing area means flat space, namely total dissipation. The transition to this state is likely to be strongly suppressed by the arguments in the last section.  Intuitively, this is because it takes time to re-absorb the huge interior, and its entropic content. Hence, the system will settle and remain on the lowest \emph{nonvanishing} eigenstate of the area for a long time.

LQG predicts the area to be quantized \cite{Rovelli1993c,Rovelli1994a}. Its eigenvalues are
\be
A_{min}=\frac{8\pi \gamma \hbar G}{c^3}\sqrt{j(j+1)}.
\ee
The lowest non vanishing eigenvalue, or ``area gap", is given by $j=\frac12$. Given the relation between  area and  mass of the horizon, this gives the mass \eqref{remmass}.

That is, LQG predicts that black holes end up as long-living particles with a mass of a few micro-grams. 

Once a remnant has attained this minimal size, it is semi-stable. It can still radiate away  its energy, and dissipate into flat space.   But this last transition 
 \begin{equation}
 |m_o,m_P\rangle \rightarrow   |0\rangle
 \end{equation}
is strongly suppressed and therefore takes a long time, at least of order $m_o^4$. 

The existence of these objects contradicts the intuition of many particle-physics or string string theorists. This is because in flat space physics, a small volume with a small energy can contain only a finite number of different states. This is what prevents the  blackbody UV catastrophe. Intuitively, to have many states we need many particles, and to have many particles with small total energy we need to have them with long wavelength, but this is not available in the volume contained within a small sphere, because this has a  small volume. This is the basis of the intuition of many particle-physics and string theorists. 

In general-relativistic physics, however, these constraints do not hold, because on a curved geometry an arbitrarily large volume can be enclosed into an arbitrarily small sphere.  Hence an arbitrarily small horizon can enclose any number of different states, limited only by the maximal size $m_o$ of the parent black hole and by the finiteness of the time during which the black hole interior has had the opportunity to grow. 

\subsection{Remnants and quantum spread of the extrinsic curvature}
\label{spread}

Let us look more closely at the physics of the remnants. Consider the geometry of a black hole in coordinates where (unlike Schwarzschild) $t=constant$ surfaces properly cut the horizon at different moments. In Lemaitre time, for instance, 
\be
ds^2=-dt^2+(dr+\sqrt{2m/r}dt)^2+r^2d\Omega^2.
\ee
The horizon $r=2m$ has area $A=16\pi m^2$. The variable conjugate to the area is the extrinsic curvature, which has a single (radial-radial) component $k=a/m^2$, where $a$ here is a numerical constant.  In quantum theory, $A$ and $k$ are conjugate variables and cannot both be sharp. We expect a Heisenberg relation of the form 
\be
\Delta A \Delta k > \hbar G. 
\ee
Classical black holes are described by semiclassical quantum states that are peaked on large values of $A$ and small values of $k$, with $\Delta A$ and $\Delta k$ satisfying the Heisenberg relation but all relative spreads being small.  Now consider an isolated black hole that radiates away all its energy, but is stuck on the minimal non-vanishing area because of the suppression of the transition to Minkowski due to its large interior.   Then the system settles on an \emph{eigenstate} of $A$.  This is the common situation of all quantum systems, left alone and free to equilibrate by radiating away energy: they settle on their minimal reachable energy eigenstate.   An eigenstate of the area will have $k$ maximally spread.  Therefore we expect  Planck scale remnants to be in states where $k$ is maximally spread.  Now consider a white hole. Its metric is easily obtained time reversing the black hole metric, that is 
\be
ds^2=-dt^2+(dr-\sqrt{2m/r}dt)^2+r^2d\Omega^2.
\ee
The area of the horizon is the same as the black hole, but $k$ has opposite sign.  An eigenstate of $A$, having $k$ maximally spread does not distinguish between a black hole and a white hole.  Hence remnants must be superpositions between the two.

\subsection{There is no information paradox} \label{InformationParadox}

We close this part of the paper by discussing the \emph{vexata questio} of the black hole information paradox, in light of the above results.  This Section follows closely \cite{Rovelli2017a} and \cite{Rovelli2019a}. 

\begin{figure}[t]
\includegraphics[width=6cm]{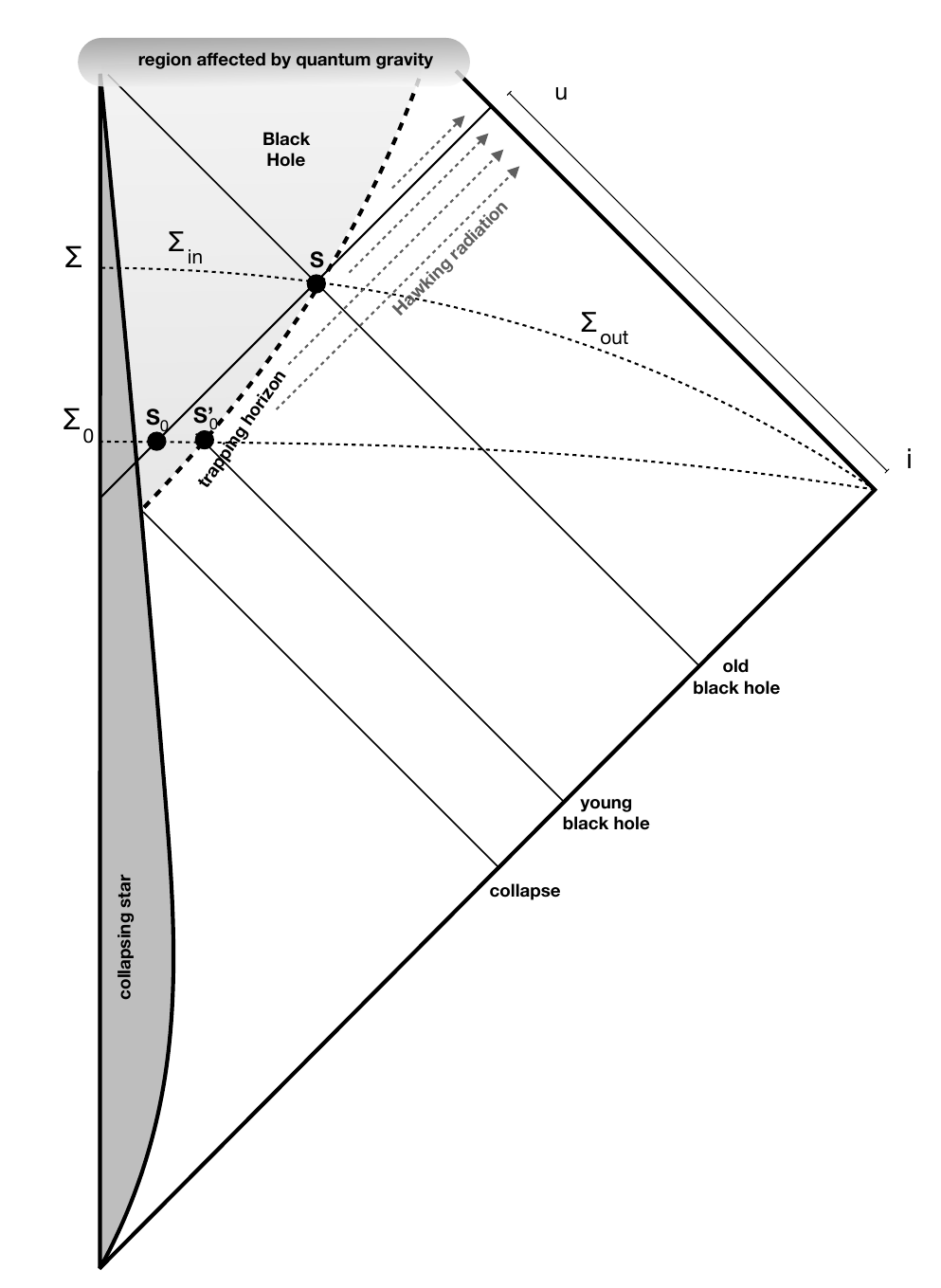}
\caption{\em The portion of spacetime relevant for the information `paradox'. }
\label{ip}
\end{figure}

The black hole information problem, as formulated by Don Page in 1993 \cite{Page1993a}, regards  the physics of the spacetime region \emph{before} the quantum gravity onset.  Figure \ref{ip} is the Carter-Penrose diagram of a  spherically symmetric spacetime geometry around a collapsing star, on which there is an evolving quantum field $\phi$. The geometry takes into account the back-reaction of the Hawking radiation of the field.   The collapsed matter distribution generates a trapped region: the black hole. The boundary of this region is the (trapping) horizon.  In the limit in which we disregard the back-reaction of the Hawking radiation, the horizon is null, but taking  back-reaction into account, it is time-like.  The notion of \emph{event} horizon is not defined, because no assumption is made about the distant future, which depends on quantum gravity and is not relevant.  Considerations on this region alone are sufficient for Page's argument for the information problem.

Here is Page's key observation. Consider a sphere $S$ on the horizon at retarded time $u$ (see Figure \ref{ip}).  The region of future null infinity preceding the time $u$ receives the Hawking radiation emitted until the horizon has reached $S$.   Consider the case where the black hole is ``old" at $S$, namely the area $A$ of $S$ is much smaller than the the initial horizon area $A_0$ at $S_0$.  The Hawking radiation arriving at future infinity around $u$ is in a mixed state.  If the initial state of the field was pure and evolution is unitary, this radiation must be correlated with something else: with what?   

There are two reasonable possibilities:  

(a) it is correlated with degrees of freedom inside the horizon; 

(b) it is correlated with degrees of freedom outside the horizon. In particular, late Hawking quanta may be correlated with early Hawking quanta. 

A part of the theoretical community has got convinced that the correct answer must be (b) because (a) is ruled out by the following argument by Don Page, based on a specific assumption. Let $N$ be the number of states of a black hole with which external degrees of freedom can be entangled at some given time and let 
\be
N_{BH} = e^{A/4}.
\ee
where $A$ is the area of the horizon at that time. Consider the following 
\be
{\rm Assumption:}\ \ \  N\sim N_{BH}.
\label{assumption}
\ee

Then, necessarily (a) is ruled out, because for sufficiently late times $A$ is small and therefore $N$ is insufficient for purifying the Hawking radiation. 

But is the assumption justified? 

The thermodynamic interaction between a black hole and its surroundings is well described by treating the black hole as a system with $N_{BH}=e^{A/4}$ (orthogonal) states, where $A$ is the horizon area. But a system can have many more degrees of freedom than those determining its  thermodynamic interactions. For instance, this is the case when some degrees of freedom are physically constrained from acting back on the exterior, as is precisely the case for a black hole. It is only for ergodic systems that the von Neumann  entangled entropy (which purifies the external state) must be equal or smaller than the thermodynamic entropy, and certainly a black hole is not ergodic. So, there is definitely no compelling reason based on known physics for assuming \eqref{assumption}.

Some theoretical hypotheses, like AdS-CFT or some String formulations elevate \eqref{assumption} into a postulate.  Juan Maldacena has called this assumption a  ``central dogma", and, like all dogmas, it can well be false.  As many dogmas, this one as well  leads to bizarre consequences, like firewalls on the horizon \cite{Almheiri2013a,Marolf2017a}.

The possibility that  $N\gg N_{BH}$ is supported by the fact that according to classical GR the interaction between a black hole and its surroundings is entirely determined by what happens in the vicinity of the horizon, not by the interior. The number $N_{BH}$ is likely to count only horizon states that (before the transition) can be distinguishable from the exterior. These are  ``surface" states. On the other hand, $N$ counts also states that can be distinguished by local observables \emph{inside} the horizon. These do not contribute to the thermodynamical entropy $S={A/4}$, but can contribute to the von Neumann entropy, which can therefore remain high even when $A$ shrinks. 

These generic considerations can be sharpened.  The fact that a blackhole can have more states than $N_{BH}$, in fact, follows from elementary considerations of causality. 

\begin{figure}[b]
\includegraphics[height=4cm]{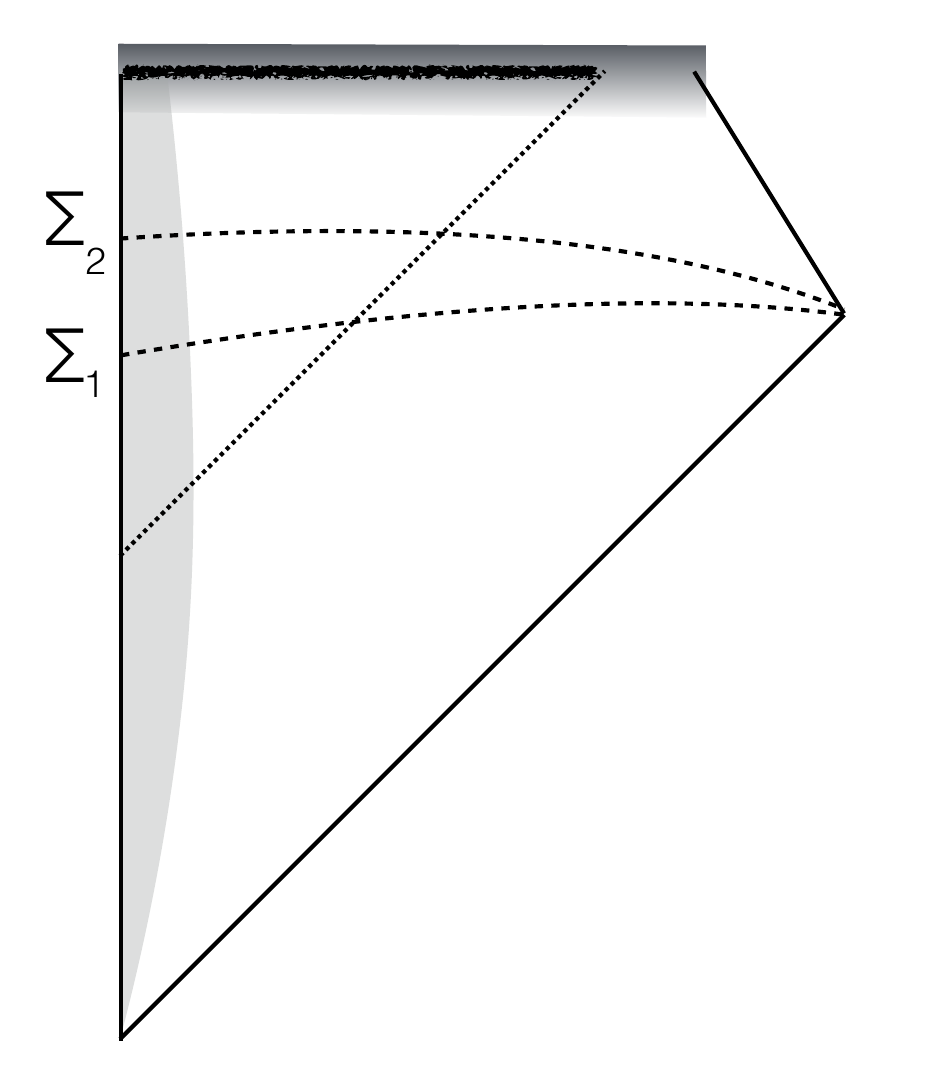}
\caption{The (lowest part) of the conformal diagram of a gravitational collapse. The clear grey region is the object, the dotted line is the horizon, the thick upper line is the singularity, the dark upper region is where quantum gravity effects may become relevant (this region plays no role in this paper.)  The two Cauchy surfaces used in the paper are the dashed lines.}
\label{f1}
\end{figure}

To show this, consider a gravitationally collapsed object and let $\Sigma_1$ be a Cauchy surface that crosses the horizon but does not hit the singularity, see Figure \ref{f1}. Let $\Sigma_2$ be a later similar Cauchy surface and $i=1,2$.  Let $A_i$ be the area of the intersection of  $\Sigma_i$ with the horizon. Assume that no positive energy falls into the horizon during the interval between the two surfaces.  Let quantum fields live on this geometry, back-reacting on it \cite{WaldCurved}. Finally, let $\Sigma_i^{in}$ be the (open) portions of $\Sigma_i$ inside the horizon. 

Care is required in specifying what is meant here by `horizon', since there are several such notions (event horizon, trapping horizon, apparent horizon, dynamical horizon...) which in this context may give tiny  differences in location.  For precision, by `horizon' we mean here the event horizon, if this exists. If it doesn't, we mean the boundary of the past of a late-time spacelike region lying outside the black hole (say outside the trapping region). With this definition, the horizon is light-like. 

Because of the back-reaction of the Hawking radiation, the area of the horizon shrinks and therefore  
\be
            A_2 < A_1. 
            \label{uno}
\ee
Now consider the evolution of the quantum fields from $\Sigma_1$ to $\Sigma_2$. We are in a region far away from the singularity and therefore (assuming the black hole is large) from high curvature.  Therefore we expect conventional quantum field theory to hold here, without  strange quantum gravity effects, at least up to high energy scales.  Since the horizon is light-like, $\Sigma_1^{in}$ is in the causal past of $\Sigma_2^{in}$.  This implies that any local observable on $\Sigma_1^{in}$ is fully determined by observables on  $\Sigma_2^{in}$.  That is, if ${\cal A}_i$ is the local algebra of observables on  $\Sigma_i^{in}$ then ${\cal A}_1$ is a subalgebra of ${\cal A}_2$:
\be
{\cal A}_1\subset {\cal A}_2.
\ee
Therefore any state on ${\cal A}_2 $ is also a state on ${\cal A}_1$ and if two such states can be distinguished by observables in ${\cal A}_1$ they certainly can be distinguished by observables in ${\cal A}_2$ as the former are included in the latter. Therefore the states that can be distinguished by ${\cal A}_1$ ---which is to say: on $\Sigma_1^{in}$--- can also be distinguished by ${\cal A}_2$ ---which is to say: on $\Sigma_2^{in}$.  Therefore the distinguishable states on $\Sigma_1^{in}$ are a subset of those in $\Sigma_2^{in}$.  How many are they? Either there is an infinite number of them, or a finite number due to some high-energy (say Planckian) cut-off. If there is an infinite number of them, then immediately the number of states distinguishable from inside the black hole is larger that $N_{NB}$, which is finite.   If there is a finite number of them, then the number $N_2$ of distinguishable states on $\Sigma_2^{in}$ must be equal or larger than the number $N_2$ of states distinguishable on $\Sigma_1^{in}$, because the second is a subset of the first. That is
\be
            N_2\ge N_1. 
            \label{due}
\ee
Comparing equations \eqref{uno} and \eqref{due} shows immediately that it is impossible that $N_i=e^{A_i/4}$, as the exponential is a monotonic function. 

This follows from only assuming the validity of quantum field theory in regions of low curvature. In other words, the ``dogma" \eqref{assumption} necessarily violates known physics in a region where we have no reasons to assume that it should fail. In our opinion, this is not good scientific method.  

The conclusion is that the number of states distinguishable from the interior of the black hole must be different from the number $N_{BH}=e^{A/4}$ of the states contributing to the Bekenstein-Hawking entropy.  Since the second is shrinking to zero with the evaporation, the first must overcome the second at some point. Therefore  in the interior of a black hole there are more possible states than $e^{A/4}$.  

The physical interpretation of the conclusion is simple: the thermal behavior of the black hole described by the Bekenstein-Hawking entropy $S=A/4$ is determined by the physics of the vicinity of the horizon. 

A vivid manifestation of the fact that in classical GR the effect of a black hole on its surroundings is independent of the black hole interior is in the numerical simulations of black hole merging and radiation emission by oscillating black holes: in writing the numerical code, it is routine to  cut away a region inside the (trapping) horizon: it is irrelevant for whatever happens outside!  This is true in classical GR, and there is no compelling reason to suppose it to fail if quantum fields are around.  

The idea of  interpreting of $S_{BH}$ as determined by the number of states of near-surface degrees of freedom, and not interior ones is of course not a new idea. It has a long history \cite{York,Zurek,Wheeler1990,THooft1990,Susskind1985,Frolov1993,Larsen1996}. See in particular \cite{Rovelli1996d}, \cite{Strominger:1997eq} in support of this idea from two different research camps, loops and strings. 

Importantly, this conclusion is \emph{not} in contrast with the various arguments leading to identify Bekenstein-Hawking entropy with a counting of states. To the opposite, evidence from it comes from the membrane paradigm \cite{Thorne1986} and from Loop Quantum Gravity \cite{Rovelli1996a,Rovelli1996d,Ashtekar:1997yu,Perez2017a}, which both show explicitly that the relevant states are surface states, but also from the string theory counting \cite{Strominger:1996sh,HorowitzStrominger}, because the counting is in a context where the relevant state space is identified with the scattering state space, which could be blind to interior observables.  For a classic discussion on different viewpoints about these alternatives, see \cite{Jacobson2005}

In conclusion, if there are more states available in a black hole than $e^{A/4}$, then the Page argument for the information loss paradox fails. States purifying the Hawking radiation are inside the hole even when the horizon shrinks. The information can later escape from the remnant emission (see below).

Notice that in popular accounts (for a recent one, see for instance \cite{Wuthrich}), three assumptions are erroneously said to be proven incompatible: (i) unitarity of the quantum evolution, (ii) equivalence principle (absence of firewalls), and (iii) quantum field theory on curved spacetimes. This is wrong.  It is only the further assumption of the dogma, that leads to problems.

\section{Elements of Phenomenology}\label{Phenomenology}

Measuring quantum gravity effects is notoriously difficult \cite{Liberati2011}.  Still, some  intriguing possibilities of observation are opened by the scenario described above.  The first is the possibility that the remnants described above are the constituents of, or contribute to, dark matter \cite{Rovelli2018f}.  In particular, remnants forming dark matter could be produced by primordial (or pre-big bang) black holes. Direct detection of these remnants has been studied in \cite{Perez2023}. A number of possible  astrophysical implications of this scenario or variants have been tentatively explored in \cite{Vidotto2016,Rovelli2017f,Barrau2017,Raccanelli2017,Vidotto2018a,Rovelli2018f,Vidotto2018,Barausse2020,barrau2021closer}.

\subsection{Dark Matter}\label{DarkMatter}

Remnants are a dark matter candidate that does not require exotic assumptions of new forces, or particles or corrections to the Einstein equations, or physics beyond the standard model. It only requires general relativity and quantum theory to hold together.  

The possibility that remnants of evaporated black holes of primordial origin could form a component of dark matter was suggested by MacGibbon \cite{J.H.MacGibbon1987} thirty years ago and has been explored by many authors \cite{Barrow1992,Carr1994,Liddle1997a,Konoplich1999, Khlopov1999,barrau2003improved,Chen2003,Chen2004,Khlopov2007a,Nozari2008, Dymnikova2015}. Since there are no strong observational constraints on this potential contribution to dark matter \cite{Carney2018a}, the weak point of the scenario has been, until now, the question of the physical nature of the remnants.  The scenario discussed in these notes has changed the  picture:  conventional physics provides a candidate for remnants.

\subsection{Direct detection} 

\begin{figure}[b]
	{\includegraphics[height=5cm]{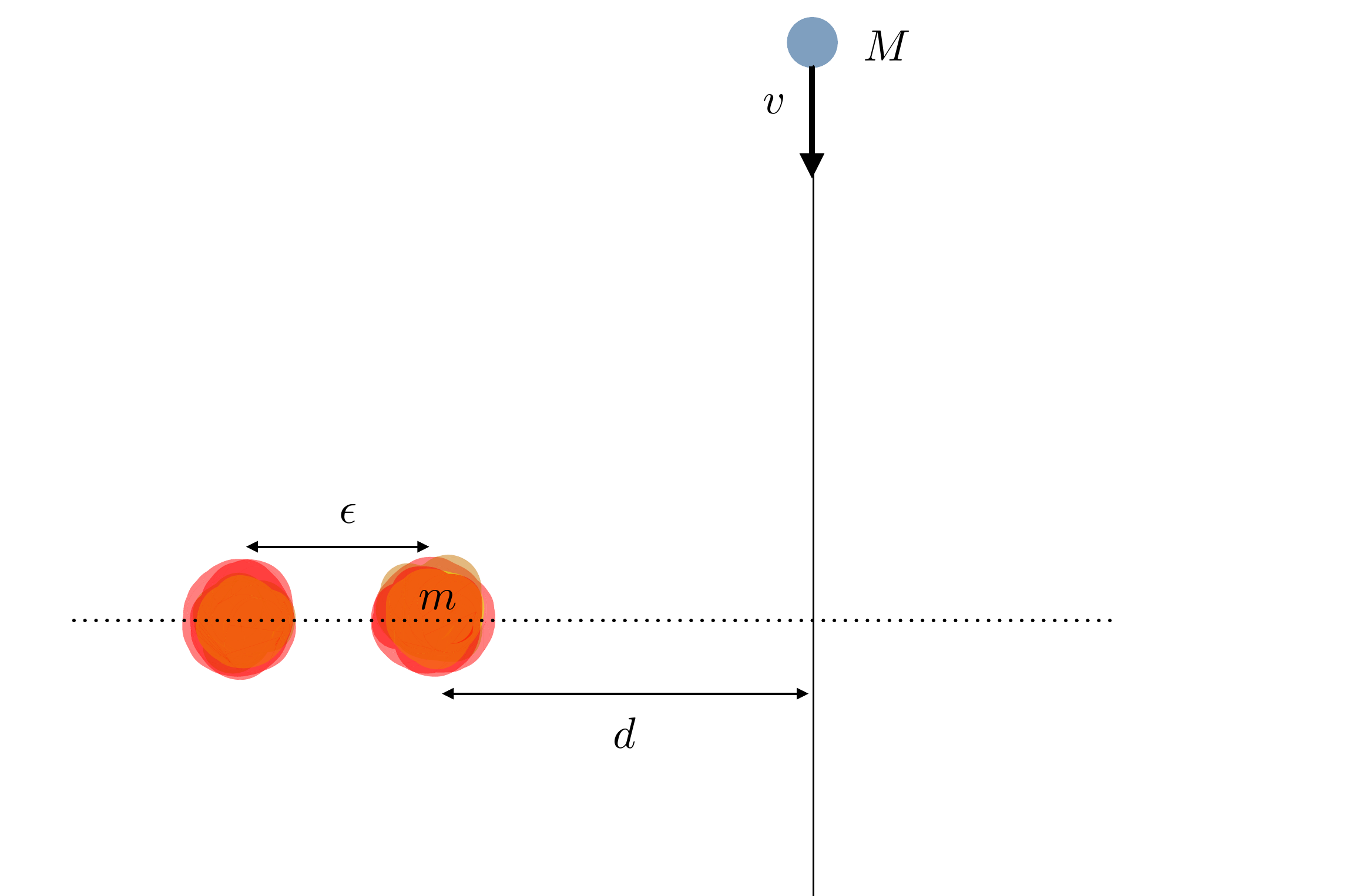}}
	\caption{A particle of mass $m$ in a superposition state with separation $\epsilon$. The DM particle passes by with velocity $v$ and a closest approach distance $d$.}.
	\label{AA-MB}
\end{figure}

A local dark matter density of the order of $0.01 M_\odot/pc^3$ corresponds to approximately one Planck-scale white hole per each $10.000 Km^3$.   These objects are presumably  moving fast with respect to our local frame, since we are rotating with the galaxy at hundreds of Km per second, while dark matter models suggest that it isn't.  This gives a very rough estimates of a few Planck scale particles per $m^2$ per year flying by us. Can we detect them?  

If the sole interaction of remnants is gravitational, they are very good dark matter candidates, but for this same reason direct detection using classical sensing is  challenging \cite{Carney2018a}, due to the extreme weakness of the gravitational interaction.  However, there may be a quantum technology that could  open a window to do the detection.  
In fact, recent developments in the area of table-top experiments involving gravity and quantum phenomena (see for instance \cite{Christodoulou2022} for up-do-date references) open the theoretical possibility of direct detection of purely-gravitationally-interacting dark matter particles.  An idealized detector where the center of detector mass is set in a superposition of locations and  a more concrete  tentative protocol, which employs Josephson junctions, have been illustrated \cite{Perez2023}. 

Figure \ref{AA-MB} illustrates the conceptual setting. The dark matter particle flies by the device, passing closer to one of the two locations in which a quantum particle is quantum split.  The different momentum transfer determines a phase difference between the two branches, that can be picked up by interferometry, or as a current change in an array of Josephson junctions.   See \cite{Perez2023} for  details.

\subsection{Cosmological implications}

Remnants could be produced by primordial black holes formed in the early universe. The lower bound of the lifetime of the remnants is of the order of or lower than $m_o^4$. If we assume that lifetime is precisely of the order of the lower bound $m_o^4$ (as mentioned, this may not be correct),  then for these objects to still be present now we need  their lifetime to be larger or equal than the Hubble time $T_H$, that is
\begin{equation}
                               m_o^4\ge T_H. 
\end{equation}
On the other hand, we expect these to be produced by evaporated black holes, therefore the lifetime of the black hole must be shorter than the Hubble time.  Therefore 
\begin{equation}
                               m_o^3  <  T_H. 
\end{equation}
This gives an estimate on the possible value of $m_0$:
\begin{equation}
                             10^{10} gr  \le m_o < 10^{15} gr.
\end{equation}
These are the masses of primordial black holes that could have given origin to dark matter present today in the form of remnants. Their Schwarzschild radius is in the range  
\begin{equation}
                             10^{-18} cm  \le R_o < 10^{-13} cm.
\end{equation}
According to primordial black hole formation theory, black holes of a given mass could have formed when their Schwarzschild radius was of the order of the horizon. Remarkably, the horizon was presumably in this range at the end of inflation, during or just after reheating.  A preliminary phenomenological analysis of this particular scenario was carried out in \cite{Barrau2014}. 

If the lifetime can be longer, constraints are less stringent. Furthermore, an intriguing possibility opens up. Remarkably, the strength of the interaction of such particles, combined with the assumption of a sufficiently hot big bang, leads to a density of these objects at decoupling whose order of magnitude is compatible with the present dark matter density \cite{Barrau2019, Amadei2022}.  

\subsection{Erebons}

An alternative possibility for the generation of remnants is that they were formed in a contracting phase before the current expanding one, in a big bounce scenario (for a review of classical and quantum bouncing cosmologies, see \cite{Brandenberger2016,Agullo2016} and references therein). 

The possibility that black holes could live across the big bounce and represent a component of dark matter has been considered in \cite{Carr2017}.  Roger Penrose has coined the name \emph{erebons}, from the Greek god of darkness Erebos, to refer to matter crossing over  from one eon to the successive one \cite{Penrose2017} in his cyclic cosmology \cite{Penrose2012}.  Large black holes evaporated before the bounce could have given rise to a population of Planck-size white hole remnants that has crossed the bounce and formed what we see today as dark matter.  For this to happen, their density should have been sufficient to balance the huge dilution in an eventual inflationary phase.  

An intriguing aspect of this scenario is the speculative idea that it might address the apparent low-entropy of the initial cosmological state \cite{Rovelli2018i}. 
This can be consistent and can represent a concrete realization of the perspectival interpretation of entropy suggested in \cite{Rovelli:2015}.  If the cosmos at the big bounce was finely dotted by white holes with large interiors, then the gravitational field was not in the very improbable low entropy homogeneous or nearly homogeneous configuration.  It was in a high entropy crumpled configuration. But being \emph{outside} all white holes, we are in a special place, and from the special perspective of this place we see the universe under a coarse grain which defines an entropy that was low in the past.

\subsection{Modeling remnants emission} 
\label{sec:emission}

Finally, remnants must themselves emit, although very softly, in order to release the information that  was trapped with the in-falling Hawking radiation and dissipate (as white holes do). 

Classically, this a very low power emission. Quantum mechanically, it is given by a weak decay probability, akin to radioactivity.  This process was tentatively modelled in  \cite{Kazemian2022a}, which we summarize here. 

Say a black hole of initial mass $m_o$ evaporates via Hawking evaporation, leaving a remnant of Planckian mass which contains an amount of information sufficient to purify its Hawking radiation, namely of order 
\be
S\sim \frac{A}4= 4\pi m_o^2
\label{S}
\ee
Here $A$ is the area of the horizon {\em at formation}.  Information is emitted in the form of radiation. Since the radiation is emitted radially, it can be modelled as a uniform one-dimensional gas of photons. Assume for simplicity  that the radiation  emitted by the surface of the remnant during the lifetime $\tau$,  is in thermal equilibrium \cite{Marolf2017a}. At the end of the remnant lifetime the radiation covers a length $L=\tau$.  The energy $E$ available for this gas is only that of the mass of the remnant, which is of the order of the Planck mass, namely unity in natural units.  
\be
E\sim 1,
\ee
while its total entropy, needed to purify the Hawking radiation is \eqref{S}. 
Entropy $S$ and energy $E$ of a 1d photon gas of temperature $T$ in a space of length $L$ are \cite{skobelev2013entropy}
\be
S=\frac{2\pi}{3}LT, \ \ \ \  E=\frac16 L T^2,
\ee
which gives 
\be
L=6m^4,\ \ \ \ T=\frac1{m^2},
\ee
and a number of quanta (say photons) emitted 
\be
N_{\gamma}\sim {m^2}.
\label{number of photons emitted}
\ee
The lifetime of a white hole would be equal to the time required for the photons to travel a distant $L$. We therefore have
\be
\tau_W\sim 6 m^4
\label{tauB}
\ee
as already mentioned, while the temperature of the white hole emission is be much lower than the initial temperature $\sim 1/m$ of the Hawking temperature of the parent black hole. 

\begin{figure}[t]
\includegraphics[width=0.5\columnwidth]{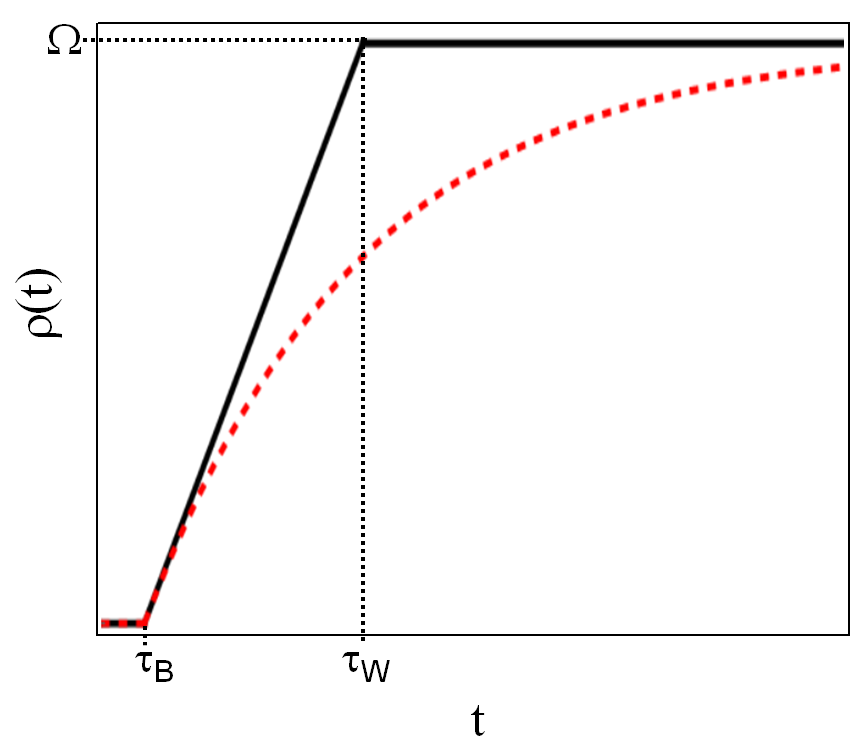}
\caption{Background white hole radiation as a function of time. The solid black line represents a classical linear emission while the dashed red line represents a quantum emission.}
\label{emissionfig}
\end{figure}

A population of black holes formed at a time $t=0$, with mass $m$ and uniformly distributed in space that evaporate around time $\tau_B\sim m^3$ as predicted by Hawking radiation theory, and survive as white hole remnants for a time $\tau_W$ as in \eqref{tauB} emits a steady radiation 
between times $\tau_B$ and $\tau_B+\tau_W$.
For $\tau_B<t<\tau_B+\tau_W$ an observer would observe that the radiation density would evolve in time as
\be
\rho(t)\left\{\begin{array}{lll}
   = 0& {\rm for} & t<m^3, \\ 
    = \big(\frac{t-\tau_B}{\tau_W-\tau_B}\big) \Omega & {\rm for} & m^3<t<6m^4, \\ 
    =\Omega & {\rm for} & t>6m^4. 
\end{array}\right.
\label{array linear emission}
\ee
In other words, the emission process is a steady (linear in time) transformation of dust into radiation, on a $m^4$ timescale. See Figure \ref{emissionfig}.

Quantum mechanics alters this picture a bit. 
A continuous energy emission as the one described above would imply a continuous decrease of the white hole horizon area, below the Planck area. But this is not permitted by LQG, because area is quantized. 

Rather, a remnant with near-Planckian mass and area must  make a single quantum leap into radiation, as in   conventional nuclear radioactivity, where emission is realized by individual quantized quanta governed by a probability distribution \cite{Martin2019}. 

In other words, in first order perturbation theory the only allowed transition is the emission of the entire Planck energy of the remnant.   

In the language of quantum field theory, this is given by a vertex between the remnant and a large number of low energy quanta. A vertex coupling a remnant to a single or a few photons, indeed, is forbidden by conservation of information (unitarity), because a few photons do not have enough degrees of freedom to match the large number of quantum numbers describing the white hole interior. Few photons  cannot carry the entire information that can be stored in the remnant. Hence the transition must be of the form  $remnant \to \gamma_1...\gamma_n $ to {\em a large number} of low energy photons: \\[1em]
\centerline{~~~~~~~~~~~~~~~~~~\includegraphics[width=0.6\columnwidth]{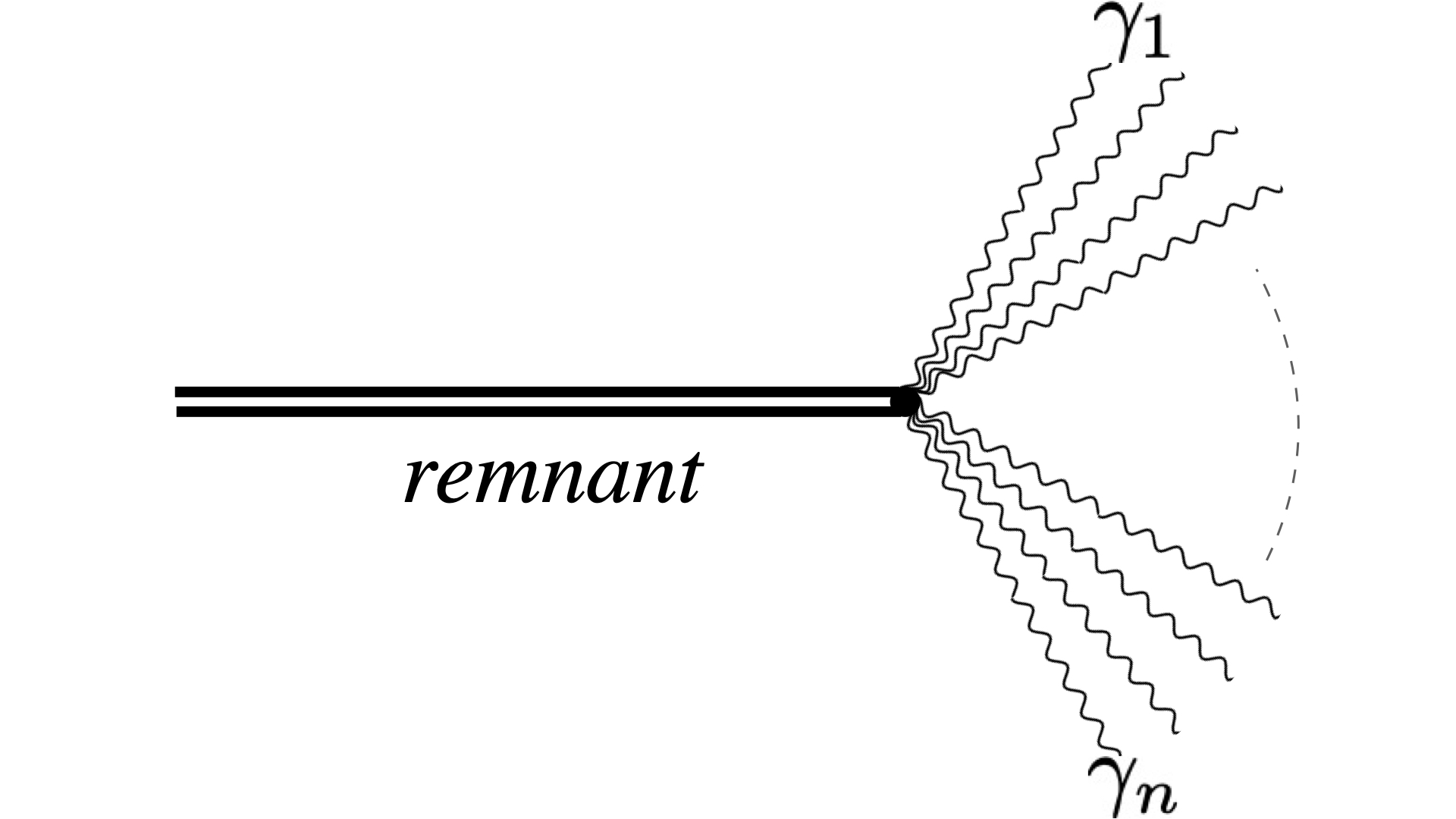}}
\\[1em]
The number of photons emitted by a single remnant is given in \eqref{number of photons emitted}. 
This conclusion is relevant in view of an old objection to the remnant scenario, that contributed to its abandonment in the Nineties. The objection was that the large number of remnant internal states would make them too easy to produce in particle physics experiments. Here we see why that conclusion was too quick.  The effective vertex responsible for a remnant production would be 
\\[1em]
\centerline{\includegraphics[width=0.6\columnwidth]{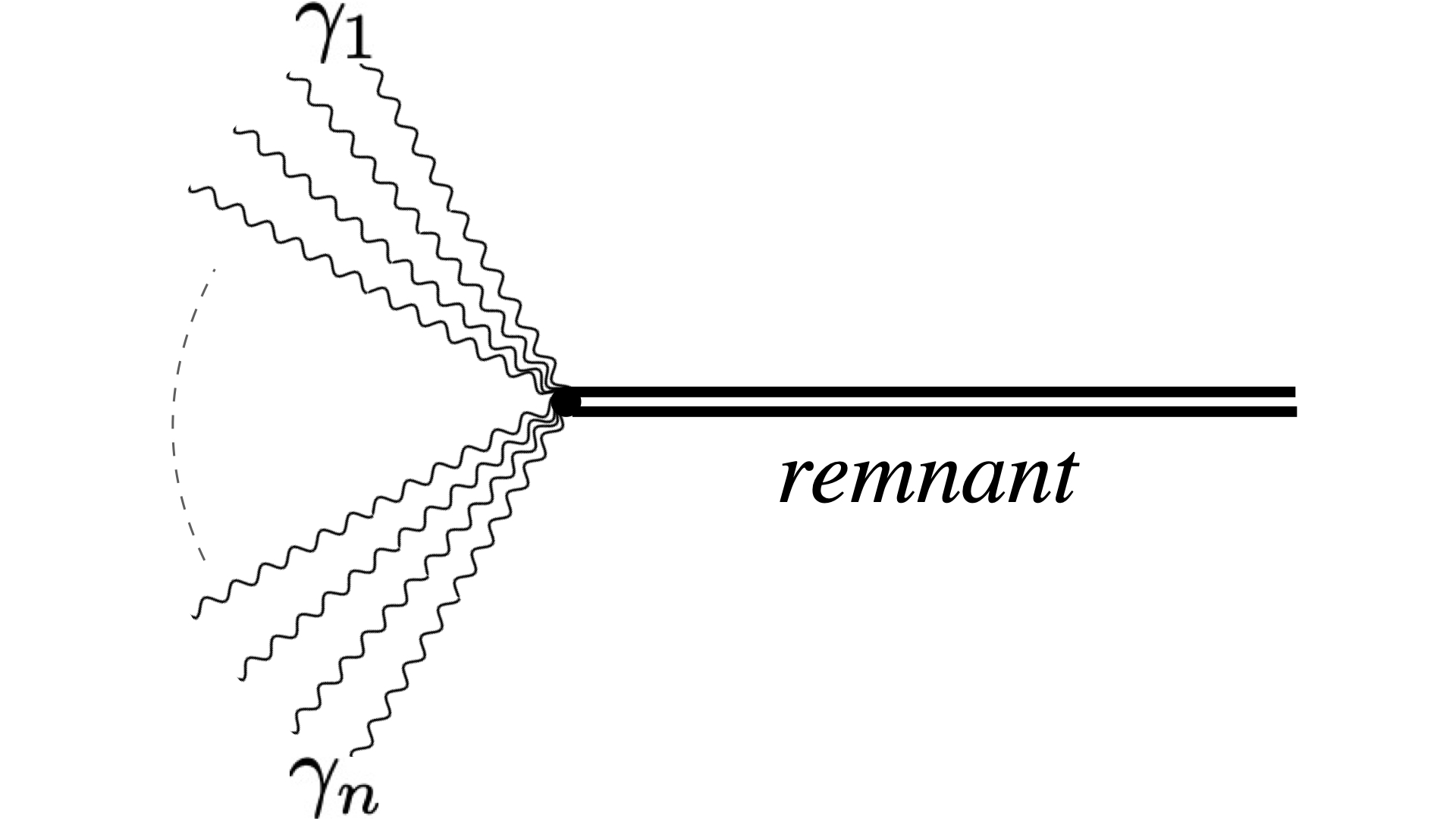}}
\\[1em]
\noindent in order to create a long living Planck size remnant.  If the number of photons was small, these could be high energy, but the remnant produced would correspond to a remnant whose parent is a black hole of Planckian size, but short lived.  The process would not be distinguishable by the standard collapse predicted by conventional quantum gravity. To produce an actual long living remnant, on the other hand, we need $m$ to be large, and hence we would need to focus {\em a large number of low energy photons}.

In the cosmological standard model, primordial black holes may have formed at reheating. To get a sense of the characteristic of the diffuse radiation that remnants may emit, one can estimate its parameter in the simplest case \cite{Kazemian2022a}.  This model is  entirely determined by a single parameter of order of unity, that can be taken to be 
\be 
x=\log_{10}(m/m_{Pl})\in[15,20].
\ee 
The results for mass, frequency and density of the emitted radiation are
\begin{eqnarray}
m &=& 10^{x-5}gr,  \\
\nu &=& 10^{-2x+32}Hz\\
\rho_{rad} &=&\sinh\bigg(\frac{10^{61}-10^{3x}}{10^{4x}-10^{3x}}\bigg) \rho_{rem}
\end{eqnarray}
This is a mass range
\be
10^{10}gr<m<10^{15}gr,
\ee
and a frequency range 
\be
10^{14}Hz>\nu>10^{4}Hz. 
\ee
Thus, a diffused radiation at frequency $\nu$ and density $\rho_{rad}$ can witness a cold dark component formed by white hole remnants, descending from primordial black holes of mass 
\be
m=10^x m_{Pl}
\ee
and with density 
\be
\rho_{rem}=\sinh^{-1}\!\!\left(\frac{1- 10^{61}\cdot 10^{-3x}}{1-6\cdot 10^{-x}}
\right) \rho_{rad}
\ee
where $x$ can be measured directly from the frequency of the diffused radiation:
\be
x=-\frac12\ \log_{10}\frac{\nu}{\nu_Pl}
\ee
In other cosmological scenarios, such as big bounce or matter bounce scenarios, remnants might also --and perhaps better-- accommodate remnants as a component of dark matter.

\section{Acknowledgments}
We thank Richard Werthamer for his accurate corrections to the manuscript.  This work was made possible through the support of the ID\# 62312 grant from the John Templeton Foundation, as part of the project \href{https://www.templeton.org/grant/the-quantum-information-structure-of-spacetime-qiss}{``The Quantum Information Structure of Spacetime'' (QISS)}.  CR acknowledges support from the Perimeter Institute for Theoretical Physics through its distinguished research chair program. FV acknowledge support from the Perimeter Institute for Theoretical Physics through its affiliation program. Research at Perimeter Institute is supported by the Government of Canada through Industry Canada and by the Province of Ontario through the Ministry of Economic Development and Innovation.
Research in FV's research group at Western University is supported by the Canada Research Chairs Program and by the Natural Science and Engineering Council of Canada (NSERC) through the Discovery Grant ``Loop Quantum Gravity: from Computation to Phenomenology."  
We acknowledge the Anishinaabek, Haudenosaunee, L\=unaap\'eewak, Attawandaron, and neutral peoples, on whose traditional lands Western University and the Perimeter Institute are located.

\bibliographystyle{utcaps}
%\bibliography{/Users/carlo/Dropbox/library}
\bibliography{library,references}
\end{document}